\documentclass[11pt]{article}
\usepackage{xcolor}
\usepackage[english]{babel}
\usepackage[T1]{fontenc}
\usepackage[utf8]{inputenc}
\usepackage{authblk}
\usepackage{mathtools}
\usepackage{mathrsfs}
\usepackage{slashed}
\usepackage{amsmath,amssymb}
\usepackage{amsfonts}
\usepackage{graphicx,color}
\usepackage{cite}
\usepackage{hyperref}
\usepackage[capitalize]{cleveref}
\usepackage{setspace}
\usepackage[left=2.5cm,right=2.5cm,top=2.5cm,bottom=2.5cm]{geometry}

\hypersetup{
colorlinks=true,
linkcolor=blue,
filecolor=magenta,
urlcolor=cyan,
}
\numberwithin{equation}{section}
\definecolor{refcol}{rgb}{0.9,0.1,0.1}
\hypersetup{colorlinks=true,linkcolor=blue,citecolor=refcol,urlcolor=cyan,linktocpage}
\allowdisplaybreaks
\newcommand{\be}{\begin{equation}}
\newcommand{\ee}{\end{equation}}
\newcommand{\bea}{\begin{eqnarray}}
\newcommand{\eea}{\end{eqnarray}}

\def\XXint#1#2#3{{\setbox0=\hbox{$#1{#2#3}{\int}$ }
\vcenter{\hbox{$#2#3$ }}\kern-.6\wd0}}

\makeatletter
\let\old@maketitle\@maketitle
\def\@maketitle{%
  \begingroup
    \let\newpage\relax 
    \begin{flushright}
      {\footnotesize\bf CQUeST-2026-0772}
    \end{flushright}
    \old@maketitle
  \endgroup
}
\makeatother

\begin{document}

	\begin{titlepage}
		\thispagestyle{empty}
		\title{ 
			{\Huge\bf The Routh of the Attractor Mechanism}
		}
		\vfill
		\author{
			{\bf Arghya Chattopadhyay${}^{1}$}\thanks{{\tt arghya.chattopadhyay@upr.edu}},
			{\bf Alessio Marrani${}^{2}$}\thanks{{\tt jazzphyzz@gmail.com}},
			{\bf Sourav Roychowdhury${}^{3}$}\thanks{{\tt srcphys@sogang.ac.kr}}\\
			\smallskip\hfill\\      
			{\small
				${}^1${\it Physics Department, University of Puerto Rico Mayag\"uez}\\
				{\it Puerto Rico 00681, USA}
				\smallskip\hfill\\        
				${}^2${\it{Centre for Mathematics and Theoretical Physics, University of Hertfordshire,} \\ 
					{\it Hatfield, Hertfordshire, AL10 9AB UK }}
				\smallskip\hfill\\ 
				${}^3${\it{Department of Physics \& Center for Quantum Spacetime, Sogang University, \\
						{\it 35 Baekbeom-ro, Mapo-gu, Seoul 04107, Republic of Korea}}}
				
			}
		}
		
		\bigskip\bigskip\bigskip\bigskip
		\vfill
		\date{		
			\begin{quote}
				\centerline{{\bf Abstract}}
				{\small
					We investigate and clarify various aspects of the effective dynamics of
Maxwell-Einstein-scalar theories in the background of static, spherically
symmetric and asymptotically flat extremal black holes in four space-time dimensions. This rigorously
places the one-dimensional effective radial dynamics governed by the
Attractor Mechanism, through the critical points of the Ferrara-Gibbons-Kallosh
effective black hole potential $V_{BH}$, into the \textit{Routhian formalism}%
, a framework which is intermediate between the Lagrange
and Hamilton ones, based on a partial Legendre transform, and especially relevant in presence of cyclic variables. We elucidate and
analyze
the interplay
of a \textit{trio} of effective functionals: the aforementioned $%
V_{BH}$, Sen's entropy functional $\mathcal{E}$, and the relevant
effective Routhian functional $\mathcal{R}$. Through their critical values at the event horizon, such functionals determine the Bekenstein-Hawking and the Wald entropy of the extremal black hole.}
			\end{quote}
            }


	
\end{titlepage}
\thispagestyle{empty}
\maketitle\vfill \eject

\tableofcontents

\newpage


\section{Introduction}

\label{sec:Introd}

The \textit{attractor mechanism} for extremal black holes is a cornerstone
of modern supergravity and string theory, providing a dynamical explanation
for the universality of extremal black hole entropy. Originally discovered
in $\mathcal{N}=2$, $D=4$ supergravity \cite{Ferrara:1995ih} and
subsequently extended to a general effective-dynamics framework by Ferrara,
Gibbons and Kallosh (FGK) \cite{Ferrara:1997tw}, the mechanism states that
scalar moduli evolving along the radial direction in the background of a
static, spherically symmetric extremal black hole approach fixed values at
the horizon which depend only on the conserved electromagnetic charges. This
ensures that the values of the scalar fields at the horizon is independent
of their asymptotic values, essentially endowing extremal solutions with a
robust \textquotedblleft no-hair\textquotedblright\ property; furthermore,
this implies that the entropy is determined only by charge-dependent duality
invariants in large classes of models.

A canonical route to the attractor behavior proceeds by symmetry-reducing
the four-dimensional Maxwell--Einstein--scalar action to an effective
one-dimensional mechanical system. In the FGK approach, this reduction
yields an effective Lagrangian for the warp factor and the scalar fields,
containing a positive-definite \textit{black hole effective potential} $V_{%
\mathrm{BH}}$ \cite{Ferrara:1997tw}. The critical points of $V_{\mathrm{BH}}$
determine the horizon values of scalars, while the Hamiltonian constraint
relates the critical value $V_{\mathrm{BH,hor}}$ to the
Bekenstein--Hawking entropy. In $\mathcal{N}=2$ theories, $V_{\mathrm{BH}}$
is naturally expressed in a manifestly symplectic covariant way, as a
consequence of the electric--magnetic duality group acting on the charge
vector; this structure can be traced back to special K\"{a}hler geometry and
its symplectic bundle formulation \cite{Ceresole:1995ca,Andrianopoli:1996cm}
(for generalization to $\mathcal{N}>2$-extended supergravities, see e.g. \cite{Andrianopoli:1997xf}).
Duality-invariant entropy formulas and the classification of attractors
solutions are deeply rooted into the resulting manifest duality covariance.

In parallel, Sen introduced the \textit{entropy function formalism} \cite%
{Sen:2005wa,Sen:2006rmg}, which is based on the universal emergence of an $%
\mathrm{AdS}_{2}$ factor of the near-horizon geometry for extremal black
holes, and computes the (Wald) entropy by extremizing a Legendre transform
of the near-horizon Lagrangian density. This method is especially powerful
for higher-derivative actions, where the relevant entropy is the
Noether-charge entropy \cite{Wald:1993nt,Iyer:1994ys}. In string theory, the
entropy function formalism has enabled precise comparisons between
macroscopic entropy computations and microscopic degeneracy counts,
beginning with the seminal $D$-brane calculation of Strominger and Vafa \cite%
{Strominger:1996sh}; later developments then established an exact counting
in supersymmetric settings \cite{Dabholkar:2004yr}.

A crucial, more recent development is the discovery that the attractor
mechanism is not restricted to supersymmetric (BPS) solutions: also \textit{%
non-supersymmetric} extremal black holes can exhibit attractor behavior
whenever $V_{\mathrm{BH}}$ admits appropriate stable critical points \cite%
{Tripathy:2005qp}. This observation gave rise to an extensive literature on
non-BPS flows, including first-order (\textquotedblleft fake
superpotential\textquotedblright ) formulations that generalize BPS gradient
flows \cite{Ceresole:2007wx}. In supergravity theories with homogeneous
symmetric scalar manifolds, group-theoretical methods based on timelike
reduction to three dimensions and nilpotent orbits lead to a systematic
construction of first-order flows for both BPS and non-BPS extremal
solutions \cite{Bossard:2009we}. Beyond the spherically symmetric case, the
interplay between attractors, multi-center solutions and stability is
generally captured by supergravity flow techniques connected to
wall-crossing and D-brane stability \cite{Denef:2000nb} (see also \cite{Ferrara:2010cw,Andrianopoli:2011gy,Ferrara:2011di,Ferrara:2012qm}).

On the geometric side, the symplectic formulation of $\mathcal{N}=2$
supergravity \cite{Ceresole:1995ca,Andrianopoli:1996cm} naturally provides a
duality-covariant arena for reformulations of the extremal black hole
attractor physics, including entropy and properties of the various classes
of critical points of $V_{\mathrm{BH}}$. An important discovery along this
venue is the so-called \textit{Freudenthal duality} \cite{Borsten:2009zy,Borsten:2012pd}: an
anti-involution on the dyonic charge vector that preserves the
(semi)classical entropy and the critical points of $V_{\mathrm{BH}}$ \cite%
{Ferrara:2011gv}.\medskip

At the same time, it is well known that dimensional reduction to an
effective radial action can be subtle in the presence of cyclic gauge
degrees of freedom and topological terms; a consistently correct treatment
of these variables is crucial to obtain the physically correct sign
structure, as well as the positivity of the resulting black hole potential
(and thus, when evaluated at the critical points of $V_{\mathrm{BH}}$, of
the Bekenstein-Hawking entropy). This observation motivates the main
conceptual goal of the present work: to place the FGK effective dynamics,
and its relation to Sen's entropy function, on a rigorous variational
footing, by introducing (and investigating the application of) the \textit{%
Routhian formalism} in supergravity.

Specifically, we will show that the one-dimensional effective action
governing the attractor flow is naturally interpreted as a \textit{Routhian}
reduction of the original four-dimensional action: namely, a \textit{partial
}Legendre transform with respect to \textit{cyclic} (gauge) variables.
Therefore, the Routhian framework, intermediate between the Lagrangian and
Hamiltonian formalisms, provides the correct variational principle in the
presence of conserved charges, and it clarifies how $V_{\mathrm{BH}}$ arises
out as a duality covariant object constructed from the symplectic
electromagnetic charge vector associated to the extremal black hole under
consideration. Within this context, we will elucidate the quite subtle
interplay among the \textit{trio} of functionals that has taken centre-stage
of the attractor literature : the FGK black hole potential $V_{\mathrm{BH}}$
\cite{Ferrara:1997tw}, Sen's entropy function $\mathcal{E}$ \cite%
{Sen:2005wa,Sen:2006rmg}, and the effective Routhian $\mathcal{R}$. Thus,
for the first time, our results provide a unified and rigorous explanation
of why the critical points of such three functionals coincide at the
horizon, and of how, in two-derivative gravity, this yields to the
equivalence between the Bekenstein--Hawking (semi)classical entropy and Wald
entropy \cite{Wald:1993nt,Iyer:1994ys}.
\bigskip

The plan of the paper is as follows.

\Cref{sec:seedattractor} deals with the effective dynamics of extremal
black holes in Maxwell-Einstein-scalar theories in four space-time
Lorentzian dimensions, and it splits into different subsections. A na\"{\i}%
ve approach to the topic (which is however part of the common lore in the
supergravity community) is discussed in \Cref{subsec:pot_gen}. This
approach, which yields to incorrect effective Lagrangians, is outclassed in
\Cref{subsec:pot_routh} by the so-called Routhian formalism, which is
intermediate between the Lagrangian and Hamiltonian frameworks, and
particularly relevant when cyclic coordinates occur: this is the rigorous
formalism to be applied to the effective dynamics of extremal black holes.
Since it may be not so well-known to the readership of this paper, in 
\Cref{sec:Resume} (in turn divided into various Subsections) we provide a 
concise yet reasonably comprehensive introduction to the main facts and
properties of such a formalism. Then, \Cref{sec:Attractors-Entropy}
deals with the attractor dynamics of scalar fields in proximity of the event
horizon of the extremal black hole, and its crucial role in the
determination of the black hole entropy, both in the Bekenstein-Hawking and
Wald approaches (whose actual equivalence is discussed in \Cref{subsec:BH-W}%
). The general relations among Sen's entropy functional $\mathcal{E}$, the
relevant Routhian functional $\mathcal{R}$, as well as the FGK effective
black hole potential $V_{BH}$, and their critical characterization at the
event horizon (determining the black hole entropy itself) are discussed in
detail in Sec. \ref{sec:efd}; in particular, after recalling the
near-horizon Bertotti-Robinson geometry of asymptotically flat extremal
black holes in \Cref{subsec:BR}, the functional $f$ (instrumental for
the definition of $\mathcal{E}$), as well as $\mathcal{E}$ itself and $\mathcal{R}$, are
respectively analyzed and their interplay elucidated in Sections \ref{f}, \ref%
{E} and \ref{R}. As to the best of our knowledge, this is the first rigorous
and comprehensive treatment of the web of 
relations among the various existing approaches to
the effective dynamics of vector and scalar fields in the background of the
extremal black holes in four dimensions; ultimately they consist into a
logical circle, which closes itself like an \textit{ouroboros} biting its
tail. After a conclusive summary and outlook given in \Cref%
{sec:Discussion}, the paper contains various Appendices, providing detailed
derivations, computations and discussions (such as Appendices \ref{app:others}, %
\ref{app:noroad} and \ref{app:genRN}), as well as a worked out example of
application of the Routhian formalism and further details about it,
respectively in Appendices \ref{App-Example-Routh} and \ref{App-Magn}.

\section{Effective dynamics of extremal black holes}

\label{sec:seedattractor}

We begin with the bosonic sector of an ungauged supergravity action in $3+1$
space-time dimensions, which is equivalent to considering the general action
of a 4-dimensional Maxwell-Einstein-scalar theory with no scalar potential
as
\begin{eqnarray}
\mathbf{S} &=&\int d^{4}x\sqrt{-g}\,L=\int d^{4}x\sqrt{-g}\left( -\frac{R}{2}%
+\frac{1}{2}g^{\mu \nu }G_{a\bar{a}}\partial _{\mu }z^{a}\partial _{\nu }%
\bar{z}^{\bar{a}}+\frac{1}{4}\mu _{\Lambda \Sigma }F_{\mu \nu }^{\Lambda
}F^{\Sigma \mu \nu }+\frac{1}{4}\nu _{\Lambda \Sigma }F_{\mu \nu }^{\Lambda
}\,{\star }F^{\Sigma \mu \nu }\right) ,  \notag \\
&&  \label{eq:theaction}
\end{eqnarray}%
where $G_{a\bar{a}}$ is the metric of the target space for the complex%
\footnote{%
With a minor loss of generality, we assume the target space to be a complex
manifold (later, we will further restrict ourselves to K\"{a}hler spaces);
this is because we mainly have in mind application to $\mathcal{N}=2$, $D=4$
supergravity. Of course, the extension to the case of real scalar manifolds
is immediate, though we will not consider it here.} scalars $z^{a}$ (with $%
a=1,...,n$) and $F^{\Lambda }$'s are the Abelian 2-form field strengths for
the $n+1$ number of one forms\footnote{%
We are using the convention ${\star }F^{\Lambda \mu \nu }={\frac{1}{2\sqrt{-g%
}}}\epsilon ^{\mu \nu \rho \sigma }F_{\rho \sigma }^{\Lambda };\,\epsilon
_{t\tau \theta \phi }=-\epsilon ^{t\tau \theta \phi }=1$.} $A^{\Lambda }$
(with $\Lambda =0,1,...,n$). $g_{\mu \nu }$ is the spacetime metric with $R$
as the Ricci scalar. The symmetric, real $\left( n+1\right) \times \left(
n+1\right) $ matrix $\mu _{\Lambda \Sigma }$ is the (negative definite)
kinetic vector matrix which enters the generalized Maxwell term $\mu
_{\Lambda \Sigma }F_{\mu \nu }^{\Lambda }F^{\Sigma \mu \nu }$; in
particular, $\mu _{\Lambda \Sigma }$ is maximal rank and thus invertible. On
the other hand, symmetric, real $\left( n+1\right) \times \left( n+1\right) $
matrix $\nu _{\Lambda \Sigma }$ enters the would-be topological term
(typical of 2-forms in $D=4$ space-time Lorentzian dimensions) $\nu
_{\Lambda \Sigma }F_{\mu \nu }^{\Lambda }\,{\star }F^{\Sigma \mu \nu }$; as
such, it needs not to be invertible, though. Notice that the (generally
non-holomorphic) dependence of $\mu _{\Lambda \Sigma }$ and $\nu _{\Lambda
\Sigma }$ on the complex scalars $z^{a}$ introduces non-trivial (and
generally non-minimal) couplings between scalars and vectors in the
framework under consideration.

The action \eqref{eq:theaction} is going to be our prototype for discussing
both Ferrara-Gibbons-Kallosh's (FGK) black hole potential formalism \cite{Ferrara:1997tw} as well as Sen's entropy function(al) formalism \cite{Sen:2007qy}. To have both the formalisms compared equally we start by considering
the following extremal, static, asymptotically flat and spherically
symmetric solution to the Einstein's equations as
\begin{equation}
ds^{2}=e^{2U(\tau )}dt^{2}-e^{-2U(\tau )}\left[ \frac{d\tau ^{2}}{\tau ^{4}}%
+\frac{1}{\tau ^{2}}\left( d\theta ^{2}+\sin ^{2}\theta d\phi ^{2}\right) %
\right] ,  \label{eq:seed_solution}
\end{equation}%
along with the fact that, as solutions to the other Euler-Lagrange equations
of motions of \eqref{eq:theaction}, only the $F_{\tau t}^{\Lambda }$ and $%
F_{\theta \phi }^{\Lambda }$ are non-zero and the scalar fields $z^{a}$ only
depend on the radial coordinate $\tau :=\frac{1}{r}$ \cite{Ferrara:1997tw, Laces_mario}.
Note that the metric \textit{Ansatz} (\ref{eq:seed_solution}) depends on a
unique function, the warp factor $U$, which is a function of $\tau $ only.
Since both FGK and Sen's formalisms have the common part of founding
themselves on an effective $1$-dimensional Lagrangian, we will start by
simplifying things as much as we can, before focusing on each formalism
individually.

Let us start and write the simplified form of the action while imposing %
\eqref{eq:seed_solution} and noting that\footnote{%
The upper dot(s) will denote differentiation with respect to $\tau $
throughout.}
\begin{align}
-\sqrt{-g}{\frac{R}{2}}& =-\sin \theta \left( \dot{U}^{2}-\ddot{U}\right) ,
\label{eq:other_terms} \\
{\frac{1}{2}}\sqrt{-g}\,g^{\mu \nu }G_{a\bar{a}}\partial _{\mu
}z^{a}\partial _{\nu }\bar{z}^{\bar{a}}& =-{\frac{1}{2}}\sin \theta \ G_{a%
\bar{a}}\dot{z}^{a}\dot{\bar{z}}^{\bar{a}}.
\end{align}%
On the other hand, the vector terms can be written as
\begin{align}
& \sqrt{-g}\frac{1}{4}g^{\mu \lambda }g^{\nu \rho }\mu _{\Lambda \Sigma
}F_{\mu \nu }^{\Lambda }F_{\lambda \rho }^{\Sigma }+\frac{1}{8}\nu _{\Lambda
\Sigma }\epsilon ^{\mu \nu \rho \sigma }F_{\mu \nu }^{\Lambda }F_{\rho
\sigma }^{\Sigma }  \notag  \label{eq:vec_terms} \\
& ={\frac{1}{2}}e^{-2U}\sin \theta \left[ -\mu _{\Lambda \Sigma }F_{\tau
t}^{\Lambda }F_{\tau t}^{\Sigma }+\mu _{\Lambda \Sigma }(\sqrt{-g}F^{\Lambda
\theta \phi })(\sqrt{-g}F^{\Sigma \theta \phi })\right] +\nu _{\Lambda
\Sigma }F_{\tau t}^{\Lambda }F_{\theta \phi }^{\Sigma }.
\end{align}%
In the expression above, we intentionally keep the second term in a format
that would suit our later treatment. The (nowadays) `standard', FGK approach
to deriving the black hole effective potential \cite{Ferrara:1997tw}
revolves around the fact that the equations of motion for the $n+1$
independent functions of $\tau $ (namely, $z^{a}$ and $U$), pertaining to %
\eqref{eq:theaction} under the metric \textit{Ansatz} (\ref{eq:seed_solution}%
), are
\begin{align}
\ddot{U}& =e^{2U}V_{BH};  \label{eq:eoms1} \\
\ddot{z}^{a}+\Gamma _{bc}^{a}\,\,\dot{z}^{b}\dot{z}^{c}& =e^{2U}G^{a\bar{b}}{%
\frac{\partial V_{BH}}{\partial \bar{z}^{\bar{b}}};}  \label{eq:eoms2} \\
\dot{U}^{2}+{\frac{1}{2}}G_{a\bar{a}}\dot{z}^{a}\dot{\bar{z}}^{\bar{a}%
}-e^{2U}V_{BH}& =0,  \label{eq:constraint}
\end{align}%
where $V_{BH}$ is the so-called `black hole effective potential', a
functional of the scalar fields $z^{a}$ which we will discuss later in some
detail. Eq. (\ref{eq:constraint}), which is a constraint\footnote{%
For a detailed derivation, one can consult the pedagogical treatment given
in the lectures \cite{Laces_mario}.}, can be derived from Einstein's
equations pertaining to the starting, 4-dimensional action (\ref%
{eq:theaction}). On the other hand, Eqs. (\ref{eq:eoms1}) and (\ref{eq:eoms2}%
) respectively are the equations of motion for the warp factor $U$ and the
complex scalar fields $z^{a}$; they follow as Euler-Lagrange equations of
the 4-dimensional action \eqref{eq:theaction} under the metric \textit{Ansatz%
} (\ref{eq:seed_solution}) and the assumptions on vectors and scalars
specified above, but they can also be regarded as the equations of motion of
the following, $1$-dimensional \textit{effective} Lagrangian \cite%
{Ferrara:1997tw}, where the role of "\textit{time}" is played by the radial
coordinate $\tau $ :
\begin{equation}
\mathcal{L}:=\dot{U}^{2}+{\frac{1}{2}}G_{a\bar{a}}\dot{z}^{a}\dot{\bar{z}}^{%
\bar{a}}+e^{2U}V_{BH},  \label{eq:normal_effective}
\end{equation}%
with the corresponding 1-dimensional effective action being given by%
\begin{equation}
\mathcal{S}=\int_{0}^{\infty }\mathcal{L}d\tau .  \label{eq:normal_action}
\end{equation}%
As mentioned above, the functional $V_{BH}$ is called the \textit{black hole
effective potential}, and will be defined and discussed further below. The
relation between $V_{BH}$ and the Bekenstein-Hawking entropy of the black
hole is fixed by the constraint \eqref{eq:constraint}, as we will derive
later for pedagogical completeness.

Before proceeding further, two observations are in order.
\begin{enumerate}
\item The l.h.s. of the equation of motion of the warp factor $U$, given by (%
\ref{eq:eoms1}), is an acceleration-like term (although keeping in mind that
$\tau $ is not the proper time, but rather the radial coordinate), but the
corresponding r.h.s. is not a force-like term, since it contains just $V_{%
\mathrm{BH}}$ and not its gradient.

\item The equations of motion for the scalar fields $z^{a}$, given by (\ref%
{eq:eoms2}), can be recast as a sort of \textit{forced geodesic equation}
with respect to the radial coordinate $\tau $, with its r.h.s. containing a
`force' $f^{a}$ given by the antiholomorphic gradient of $V_{\mathrm{BH}}$
(namely, $f^{a}:=G^{a\bar{b}}\frac{\partial V_{\mathrm{BH}}}{\partial z^{%
\bar{b}}}$), and with a $\tau $-dependent `mass' $\breve{m}\left( \tau
\right) :=e^{-2U(\tau )}$. In fact, (\ref{eq:eoms2}) can be rewritten as%
\begin{eqnarray}
\breve{m}\left( \tau \right) \frac{D\dot{z}^{a}}{d\tau } &=&f^{a}, \\
\frac{D}{d\tau } &:=&\dot{z}^{a}\bigtriangledown _{a},
\end{eqnarray}%
where $\bigtriangledown _{a}$ is the covariant derivative in the
(Riemannian) scalar manifold (with the standard Christoffel connection).
Concerning the near-horizon behaviour of $\breve{m}\left( \tau \right) $,
cf. Eq. (\ref{eq:at_horizon}) further below.
\end{enumerate}

\noindent We will now focus onto deriving
the effective Lagrangian $\mathcal{L}$ (\ref{eq:normal_effective}), first in
a na\"{\i}ve way, and then investigating its natural appearance (and thus,
explanation) within the Routhian formalism.

\subsection{A na\"{\i}ve approach}\label{subsec:pot_gen}

One could na\"{\i}vely expect to derive the $1$-dimensional effective action
$\mathcal{S}$ \eqref{eq:normal_action} from the starting, $4$-dimensional
action \eqref{eq:theaction} by using the equations of motion with the
relations \eqref{eq:other_terms} and \eqref{eq:vec_terms}; this is the
procedure carried out by Kallosh, Sivanandam and Soroush in \cite%
{Kallosh:2006bt} (cf. App. A therein, and the brief review we provide in
App. \ref{app:noroad}) . However, this procedure holds at the cost of
introducing \textit{unphysical}, purely imaginary dual 2-form field
strengths for the Abelian 2-form field strengths $F^{\Lambda }$. Actually,
the derivation of \eqref{eq:normal_effective} needs more careful
computations, which, as we will see below, will also naturally determine the
connection to Sen's formalism. Before moving on to such a careful and
thorough derivation, we now spend a few lines in order to show the issues
that one would encounter while following the (apprently and) supposedly
correct route.

One can start by considering \eqref{eq:other_terms} and \eqref{eq:vec_terms}
all together, and drop the total derivative term\footnote{%
Of course, this holds when assuming trivial topology of the considered
space-time, or at least some minimal conditions for the vanishing of the
total derivative term. We will not further deal with such possible issues in
the present work.} to yield an effective Lagrangian%
\begin{equation}\label{eq:lag_eff}
\mathcal{L}=\sin \theta \left( \mathcal{L}_{\text{scalar}}+\mathcal{L}_{%
\text{vector}}\right) ,
\end{equation}%
where
\begin{align}
\mathcal{L}_{\text{scalar}}=& -\dot{U}^{2}-\frac{1}{2}G_{a\bar{a}}\dot{z}^{a}%
\dot{\bar{z}}^{\bar{a}};  \notag \\
& \\
\mathcal{L}_{\text{vector}}=& \frac{1}{2}e^{-2U}\left[ -\mu _{\Lambda \Sigma
}F_{\tau t}^{\Lambda }F_{\tau t}^{\Sigma }+\mu _{\Lambda \Sigma }\left(
\sqrt{-g}F^{\Lambda \theta \phi }\right) \left( \sqrt{-g}F^{\Sigma \theta
\phi }\right) \right]  \notag \\
& +e^{-2U}\ \nu _{\Lambda \Sigma }\epsilon ^{\tau t\theta \phi }F_{\tau
t}^{\Lambda }\left( \sqrt{-g}F^{\Sigma \theta \phi }\right) \ .
\end{align}%
One can then proceed with further simplifications by defining the field
strengths $F^{\Lambda }$ and their dual $\mathcal{G}_{\Lambda }$ in term of
the electric and magnetic potentials, respectively denoted by $\psi
^{\Lambda }$ and $\chi _{\Lambda }$, as
\begin{align}
F_{\tau t}^{\Lambda }=\partial _{\tau }\psi ^{\Lambda },\,\,& \mathcal{G}%
_{\Lambda \tau t}=\partial _{\tau }\chi _{\Lambda };  \label{eq:field_pot} \\
\mathcal{G}_{\Lambda \mu \nu }=-\epsilon _{\mu \nu \rho \sigma }\frac{\delta
\mathcal{L}_{\text{vector}}}{\delta F_{\rho \sigma }^{\Lambda }}& =-\mu
_{\Lambda \Sigma }\star F_{\mu \nu }^{\Sigma }+\nu _{\Sigma \Lambda }F_{\mu
\nu }^{\Sigma },  \label{eq:field_pot_2}
\end{align}%
which leads to the relation
\begin{equation}
\sqrt{-g}F^{\Lambda \theta \phi }=\left( \mu ^{-1}\right) ^{\Lambda \Sigma
}\partial _{\tau }\chi _{\Sigma }-\left( \mu ^{-1}\nu \right) _{\Sigma
}^{\Lambda }\partial _{\tau }\psi ^{\Sigma }\ .  \label{eq:field_pot2}
\end{equation}%
In terms of the potentials $\psi ^{\Lambda }$ and $\chi _{\Lambda }$, the
first term of $\mathcal{L}_{\text{vector}}$ can thus be written as
\begin{equation}
\frac{1}{2}e^{-2U}\sin \theta \ \left(
\begin{array}{cc}
\partial _{\tau }\psi ^{\Lambda }\ ~ & \partial _{\tau }\chi _{\Lambda }%
\end{array}%
\right)
\begin{pmatrix}
\left( -\mu +\nu \mu ^{-1}\nu \right) _{\Lambda \Sigma } & -\left( \nu \mu
^{-1}\right) _{\Lambda }^{\Sigma } \\
&  \\
-\left( \mu ^{-1}\nu \right) _{\Sigma }^{\Lambda } & \left( \mu ^{-1}\right)
^{\Lambda \Sigma } \\
&
\end{pmatrix}%
\begin{pmatrix}
\partial _{\tau }\psi ^{\Sigma } \\
\\
\partial _{\tau }\chi _{\Sigma } \\
\end{pmatrix}%
\ .  \label{eq:vt_first}
\end{equation}%
Similarly, in terms of the potentials the second term of $\mathcal{L}_{\text{%
vector}}$ becomes
\begin{equation}
e^{-2U}\sin \theta \left[ \left( \nu \mu ^{-1}\right) _{\Lambda }^{\Sigma
}\partial _{\tau }\psi ^{\Lambda }\partial _{\tau }\chi _{\Sigma }-\left(
\nu \mu ^{-1}\nu \right) _{\Lambda \Sigma }\partial _{\tau }\psi ^{\Lambda
}\partial _{\tau }\psi ^{\Sigma }\right] \ .  \label{eq:vt_second}
\end{equation}%
Consequently, one can rewrite the Lagrangian $\mathcal{L}$ as
\begin{eqnarray}
\mathcal{L} &=&-\sin \theta \left( \dot{U}^{2}+\frac{1}{2}G_{a\bar{a}}\dot{z}%
^{a}\dot{\bar{z}}^{\bar{a}}\right)  \notag \\
&&+\frac{1}{2}e^{-2U}\sin \theta \ \left(
\begin{array}{cc}
\partial _{\tau }\psi ^{\Lambda }\ ~ & \partial _{\tau }\chi _{\Lambda }%
\end{array}%
\right)
\begin{pmatrix}
\left( -\mu -\nu \mu ^{-1}\nu \right) _{\Lambda \Sigma } & \left( \nu \mu
^{-1}\right) _{\Lambda }^{\Sigma } \\
&  \\
-\left( \mu ^{-1}\nu \right) _{\Sigma }^{\Lambda } & \left( \mu ^{-1}\right)
^{\Lambda \Sigma } \\
&
\end{pmatrix}%
\begin{pmatrix}
\partial _{\tau }\psi ^{\Sigma } \\
\\
\partial _{\tau }\chi _{\Sigma } \\
\end{pmatrix}%
\ .  \label{eq:eff_lag_4d}
\end{eqnarray}%
One can now observe that the angular part of $\mathcal{L}$ completely
decouples from the radial part; therefore, upon integrating out the angular
part, $\mathcal{L}$ acquires the form\footnote{%
Up to an overall $4\pi $ factor from integrating $\iint d\theta d\phi \sin
\theta $.}
\begin{eqnarray}
\mathcal{L} &=&-\dot{U}^{2}-\frac{1}{2}G_{a\bar{a}}\dot{z}^{a}\dot{\bar{z}}^{%
\bar{a}}  \notag \\
&&+\frac{1}{2}e^{-2U}\left(
\begin{array}{cc}
\partial _{\tau }\psi ^{\Lambda }\ ~ & \partial _{\tau }\chi _{\Lambda }%
\end{array}%
\right)
\begin{pmatrix}
\left( -\mu -\nu \mu ^{-1}\nu \right) _{\Lambda \Sigma } & \left( \nu \mu
^{-1}\right) _{\Lambda }^{\Sigma } \\
&  \\
-\left( \mu ^{-1}\nu \right) _{\Sigma }^{\Lambda } & \left( \mu ^{-1}\right)
^{\Lambda \Sigma } \\
&
\end{pmatrix}%
\begin{pmatrix}
\partial _{\tau }\psi ^{\Sigma } \\
\\
\partial _{\tau }\chi _{\Sigma } \\
\end{pmatrix}%
\ .  \label{eq:eff_lag_1d}
\end{eqnarray}%
The \emph{mechanical} (1-dimensional, effective) Lagrangian $\mathcal{L}$
given by \eqref{eq:eff_lag_1d} exhibits the quite peculiar feature that its
dynamics can be described in terms of the radial coordinate $\tau $ rather
than the time, which governs the dynamics in the standard Lagrngian
formalism. The overall $U(1)^{n+1}$ gauge invariance of the starting,
4-dimensional Lagrangian $L$ (\ref{eq:theaction}) implies $\mathcal{L}$ to
be independent of the 1-form Abelian electric-magnetic potentials $\psi
^{\Lambda }$ and $\chi _{\Lambda }$, which are therefore \textit{cyclic}
fields of $\mathcal{L}$ itself. Thus, the relevant version of the Noether
theorem implies the existence of $2(n+1)$ conserved charges of the $1$%
-dimensional, effective radial dynamics described by $\mathcal{L}$. Their
explicit expressions read
\begin{eqnarray}
\tilde{p}^{\Gamma } &:=&\frac{\delta \mathcal{L}}{\delta (\partial _{\tau
}\chi _{\Gamma })}=\frac{1}{2}e^{-2U}\left[ 2\left( \mu ^{-1}\right)
^{\Lambda \Gamma }\partial _{\tau }\chi _{\Lambda }-\left( \mu ^{-1}\nu
\right) _{\Sigma }^{\Gamma }\partial _{\tau }\psi ^{\Sigma }+\left( \nu \mu
^{-1}\right) _{\Lambda }^{\Gamma }\partial _{\tau }\psi ^{\Lambda }\right] \
,  \notag \\
\tilde{q}_{\Gamma }&:=&\frac{\delta \mathcal{L}}{\delta (\partial _{\tau
}\psi ^{\Gamma })}=\frac{1}{2}e^{-2U}\left[ 2\left( -\mu -\nu \mu ^{-1}\nu
\right) _{\Gamma \Sigma }\partial _{\tau }\psi ^{\Sigma }-\left( \mu
^{-1}\nu \right) _{\Gamma }^{\Lambda }\partial _{\tau }\chi _{\Lambda
}+\left( \nu \mu ^{-1}\right) _{\Gamma }^{\Sigma }\partial _{\tau }\chi
_{\Sigma }\right] \ .  \notag \\
&&  \label{eq:eff_lag_charges}
\end{eqnarray}%
The fact that these are integrals of the (radial) motion follows from the
fact that $\psi ^{\Gamma }$ and $\chi _{\Gamma }$ are cyclic fields of $%
\mathcal{L}$ : indeed, the Euler-Lagrange equations for $\mathcal{L}$ imply%
\begin{eqnarray}
\frac{d\tilde{p}^{\Gamma }}{d\tau } &=&\frac{d}{d\tau }\frac{\delta \mathcal{%
L}}{\delta (\partial _{\tau }\chi _{\Gamma })}=\frac{\delta \mathcal{L}}{%
\delta \chi _{\Gamma }}=0; \\
\frac{d\tilde{q}_{\Gamma }}{d\tau } &=&\frac{d}{d\tau }\frac{\delta \mathcal{%
L}}{\delta (\partial _{\tau }\psi ^{\Gamma })}=\frac{\delta \mathcal{L}}{%
\delta \psi ^{\Gamma }}=0.
\end{eqnarray}%
A concise version of (\ref{eq:eff_lag_charges}) and its inversion
resplctively read
\begin{equation}
\begin{pmatrix}
\tilde{q}_{\Gamma } \\
\\
\tilde{p}^{\Gamma } \\
\end{pmatrix}%
=2e^{-2U}\mathcal{K}%
\begin{pmatrix}
\partial _{\tau }\psi ^{\Gamma } \\
\\
\partial _{\tau }\chi _{\Gamma } \\
\end{pmatrix}%
\Leftrightarrow
\begin{pmatrix}
\partial _{\tau }\psi ^{\Gamma } \\
\\
\partial _{\tau }\chi _{\Gamma } \\
\end{pmatrix}%
=\frac{1}{2}e^{2U}\mathcal{K}^{-1}%
\begin{pmatrix}
\tilde{q}_{\Gamma } \\
\\
\tilde{p}^{\Gamma } \\
\end{pmatrix}%
\ ,  \label{eq:pot_to_charge}
\end{equation}%
where the $2\left( n+1\right) \times 2\left( n+1\right) $ real symmetric
matrix $\mathcal{K}$ is defined as follows :
\begin{equation*}
\mathcal{K}:=%
\begin{pmatrix}
\left( -\mu -\nu \mu ^{-1}\nu \right) _{\Lambda \Sigma } & \left( \nu \mu
^{-1}\right) _{\Lambda }^{\Sigma } \\
&  \\
-\left( \mu ^{-1}\nu \right) _{\Sigma }^{\Lambda } & \left( \mu ^{-1}\right)
^{\Lambda \Sigma } \\
&
\end{pmatrix}%
\ .
\end{equation*}
Finally, one can write \footnote{%
We take the transpose of \eqref{eq:pot_to_charge} to write
\begin{equation*}
\left( \partial _{\tau }\psi ^{\Gamma }\ \partial _{\tau }\chi _{\Gamma
}\right) =\frac{1}{2}e^{2U}\left( \tilde{q}_{\Gamma }\ \tilde{p}^{\Gamma
}\right) \left( \mathcal{\tilde{M}}^{-1}\right) ^{T}\ .
\end{equation*}%
} an effective one dimensional lagrangian $\mathcal{L}_{\text{eff}}$\footnote{The minus sign is absorbed while defining $\mathcal{L}_{\text{eff}}$ to match the convention of \cite{Ferrara:1997tw} apart from the sign convention for the matrices $\mu_{\Lambda\Sigma}$ and $\nu_{\Lambda\Sigma}$} as
\begin{equation}
\mathcal{L_{\text{eff}}}=-\mathcal{L}=\dot{U}^{2}+\frac{1}{2}G_{a\bar{a}}\dot{z}^{a}\dot{\bar{z}}^{%
\bar{a}}+e^{2U}\mathcal{V}_{BH},  \label{eq:eff_laf_full}
\end{equation}%
where the functional $\mathcal{V}_{BH}$ is defined as follows :
\begin{equation}\label{eq:wrong_pot}
\mathcal{V}_{BH}:=\frac{1}{4}\left(
\begin{array}{cc}
\tilde{p}^{\Lambda }~ & \tilde{q}_{\Lambda }%
\end{array}%
\ \right)
\begin{pmatrix}
\left( \mu +\nu \mu ^{-1}\nu \right) _{\Lambda \Sigma } & \left( \nu \mu
^{-1}\right) _{\Lambda }^{\Sigma } \\
&  \\
-\left( \mu ^{-1}\nu \right) _{\Sigma }^{\Lambda } & -\left( \mu ^{-1}\right)
^{\Lambda \Sigma } \\
&
\end{pmatrix}%
\begin{pmatrix}
\tilde{p}^{\Sigma } \\
\\
\tilde{q}_{\Sigma } \\
\end{pmatrix}%
\ .
\end{equation}%
Unexpectedly (if one had to follow the common \textit{folklore}), the
functional $\mathcal{V}_{BH}$ occurring in (\ref{eq:eff_laf_full}) is
different from the (still-to-be-defined) usual black hole effective
potential $V_{BH}$ occurring in the $1$-dimensional effective action $%
\mathcal{S}$ given by (\ref{eq:normal_effective}); indeed, $\mathcal{V}_{BH}$
differs from by signs in front of the vector coupling terms in its diagonal
as well as in its off-diagonal entries. This results in $\mathcal{V}_{BH}$
being not positive definite, and thus in \eqref{eq:eff_laf_full} not being
the correct effective $1$-dimensional Lagrangian to be chosen. In order to
resolve the sign issue in \eqref{eq:eff_laf_full}, and thus switch from $%
\mathcal{V}_{BH}$ to $V_{BH}$, we will proceed to prescribe the following
approach\footnote{Actually, one can try to use a Routhian functional directly starting from \eqref{eq:eff_lag_1d} in multiple different ways, but,
as shown in \cref{app:others}, all of them will produce incorrect black hole potentials.}.

\subsection{$\mathcal{R}$ and the Routhian approach}

\label{subsec:pot_routh} We will start from \eqref{eq:lag_eff}, and first
use the relation%
\begin{equation}
F_{\tau t}^{\Lambda }=\partial _{\tau }\psi ^{\Lambda },
\end{equation}
namely the first of (\ref{eq:field_pot}), \textit{without} introducing the
dual field strength $\mathcal{G}_{\Lambda }$ : in this case, only $F_{\tau
t}^{\Lambda }$, and not also $F^{\Lambda |\theta \phi }$, works as a \textit{%
momentum-like}\footnote{%
Since we are dealing with radial evolution, this is not exactly momentum.}
variable. Therefore, in this scenario \textit{there is only one cyclic
coordinate} $\psi ^{\Lambda }$, which produces the conserved charge
\begin{equation*}
Q_{\Lambda }:={\frac{\delta \mathcal{L}}{\delta (\partial _{\tau }\psi
^{\Lambda })}}.
\end{equation*}

Our prescription to reproduce \eqref{eq:normal_effective} from %
\eqref{eq:lag_eff} starts by first writing down \eqref{eq:lag_eff} in terms
of the conserved charge $Q_{\Lambda }$, and then perform the Legendre
transform of \eqref{eq:lag_eff} with respect to the cyclic coordinate $\psi^\Lambda$. Since we are not performing any Legendre transform with respect to any of the rest "\textit{momenta-like}" variables, we will
finally get not Hamiltonian functional, but something else, which goes under
the name of \textit{Routhian functional}; for a brief review of the Routhian formalism, see the
next Section. The explicit choice of Routhian we make here reads\footnote{%
We use $\dot{\psi}^{\Lambda }\equiv \partial _{\tau }\psi ^{\Lambda }$.}
\begin{equation}
\mathcal{R}:=\dot{\psi}^{\Lambda }{%
\frac{\delta \mathcal{L}}{\delta \dot{\psi}^{\Lambda }}}-\mathcal{L}\ .
\label{eq:Routh_def}
\end{equation}%
By virtue of \eqref{eq:lag_eff}, this can be recast as follows :
\begin{equation*}
\mathcal{R}=\sin \theta \left[ \dot{U}^{2}+\frac{1}{2}G_{a\bar{a}}\dot{z}^{a}%
\dot{\bar{z}}^{\bar{a}}-\frac{1}{2}e^{-2U}\left( \mu _{\Lambda \Sigma
}F_{t\tau }^{\Lambda }F_{t\tau }^{\Sigma }+\mu _{\Lambda \Sigma }\left(
\sqrt{-g}F^{\Lambda \theta \phi }\right) \left( \sqrt{-g}F^{\Sigma \theta
\phi }\right) \right) \right] \ .
\end{equation*}

Next, under the same definition of dual field strength and the magnetic
potential as in \eqref{eq:field_pot}-\eqref{eq:field_pot2}, $\mathcal{R}$
can be rewritten in terms of the electric and magnetic potential as\footnote{%
Again, up to an overall factor of $4\pi $, potentially originated by the
angular integration.}
\begin{eqnarray}
\mathcal{R} &=&\dot{U}^{2}+\frac{1}{2}G_{a\bar{a}}\dot{z}^{a}\dot{\bar{z}}^{%
\bar{a}}  \notag \\
&&-\frac{1}{2}e^{-2U}\left(
\begin{array}{cc}
\partial _{\tau }\psi ^{\Lambda }\ ~ & \partial _{\tau }\chi _{\Lambda }%
\end{array}%
\right)
\begin{pmatrix}
\left( \mu +\nu \mu ^{-1}\nu \right) _{\Lambda \Sigma } & -\left( \nu \mu
^{-1}\right) _{\Lambda }^{\Sigma } \\
&  \\
-\left( \mu ^{-1}\nu \right) _{\Sigma }^{\Lambda } & \left( \mu ^{-1}\right)
^{\Lambda \Sigma } \\
&
\end{pmatrix}%
\begin{pmatrix}
\partial _{\tau }\psi ^{\Sigma } \\
\\
\partial _{\tau }\chi _{\Sigma } \\
\end{pmatrix}%
,  \notag \\
&&  \label{ro}
\end{eqnarray}%
where we have integrated out the angular coordinates.

Again, we observe that $\mathcal{R}$ (\ref{ro}) is manifestly independent of
$\psi ^{\Lambda }$ and $\chi _{\Lambda }$ and only their derivatives appear.
Therefore, such two sets of cyclic coordinates correspondingly give rise to
two sets of conserved charges (integrals of the radial dynamics, to be
contrasted with their tilded counterparts given by (\ref{eq:eff_lag_charges}%
)), which are constants for the radial dynamics governed by $\mathcal{R}$ :
\begin{eqnarray}
p^{\Gamma } &:=&\frac{\delta \mathcal{R}}{\delta (\partial _{\tau }\chi
_{\Gamma })}=-\frac{1}{2}e^{-2U}\left[ 2\left( \mu ^{-1}\right) ^{\Lambda
\Gamma }\partial _{\tau }\chi _{\Lambda }-\left( \mu ^{-1}\nu \right)
_{\Sigma }^{\Gamma }\partial _{\tau }\psi ^{\Sigma }-\left( \nu \mu
^{-1}\right) _{\Lambda }^{\Gamma }\partial _{\tau }\psi ^{\Lambda }\right] \
,  \notag \\
q_{\Gamma }&:=&\frac{\delta \mathcal{R}}{\delta (\partial _{\tau }\psi
^{\Gamma })}=-\frac{1}{2}e^{-2U}\left[ 2\left( \mu +\nu \mu ^{-1}\nu \right)
_{\Gamma \Sigma }\partial _{\tau }\psi ^{\Sigma }-\left( \mu ^{-1}\nu
\right) _{\Gamma }^{\Lambda }\partial _{\tau }\chi _{\Lambda }-\left( \nu
\mu ^{-1}\right) _{\Gamma }^{\Sigma }\partial _{\tau }\chi _{\Sigma }\right]
\ .  \notag \\
&&  \label{eq:Routh_consts}
\end{eqnarray}%
These are called magnetic and electric charges of the extremal black hole
under consideration, respectively. The fact that they are integrals of the
(radial) motion follows from the fact that $\psi ^{\Gamma }$ and $\chi
_{\Gamma }$ are cyclic fields of $\mathcal{R}$ : indeed, the Euler-Lagrange
equations for $\mathcal{R}$ imply%
\begin{eqnarray}
\frac{dp^{\Gamma }}{d\tau } &=&\frac{d}{d\tau }\frac{\delta \mathcal{R}}{%
\delta (\partial _{\tau }\chi _{\Gamma })}=\frac{\delta \mathcal{R}}{\delta
\chi _{\Gamma }}=0; \\
\frac{dq_{\Gamma }}{d\tau } &=&\frac{d}{d\tau }\frac{\delta \mathcal{R}}{%
\delta (\partial _{\tau }\psi ^{\Gamma })}=\frac{\delta \mathcal{R}}{\delta
\psi ^{\Gamma }}=0.
\end{eqnarray}%
Then, a little mathematical manipulation will reveal that $\mathcal{R}$
enjoys the concise form
\begin{equation}
\mathcal{R}=\dot{U}^{2}+{\frac{1}{2}}G_{a\bar{a}}\dot{z}^{a}\dot{\bar{z}}^{%
\bar{a}}+e^{2U}V_{BH},  \label{eq:finalrouth}
\end{equation}%
with the \textit{black hole effective potential} $V_{BH}$ defined as
\begin{equation}
V_{BH}:=-{\frac{1}{2}}\mathcal{Q}^{T}\mathcal{M}\,\mathcal{Q},  \label{vbh}
\end{equation}%
where $\mathcal{Q}$ is the symplectic vector of magnetic and electric
charges of the extremal black hole,%
\begin{equation}
\mathcal{Q}:=%
\begin{pmatrix}
p^{\Lambda } & q_{\Lambda }%
\end{pmatrix}%
^{T},
\end{equation}%
and $\mathcal{M}$ is a $2(n+1)\times 2(n+1)$ real, symmetric, negative
definite and symplectic matrix \cite{Laces_mario, Ferrara:1997tw},%
\begin{equation}
\mathcal{M}:=%
\begin{pmatrix}
\left( \mu +\nu \mu ^{-1}\nu \right) _{\Lambda \Sigma } & \left( \nu \mu
^{-1}\right) _{\Lambda }^{\Sigma } \\
&  \\
\left( \mu ^{-1}\nu \right) _{\Sigma }^{\Lambda } & \left( \mu ^{-1}\right)
^{\Lambda \Sigma } \\
&
\end{pmatrix}%
.
\end{equation}%
The negative definiteness of $\mathcal{M}$ (which ultimately follows from
the negative definiteness of $\mu _{\Lambda \Sigma }$) implies $V_{BH}$ to
be positive definite :
\begin{equation}
V_{BH}\geqslant 0,
\end{equation}%
thus ensuring that one can view $\mathcal{R}$ as an effective, $1$%
-dimensional functional describing the dynamics of a system in a positive
definite potential, with the role of time played by the radial coordinate $%
\tau $.

The result (\ref{eq:finalrouth}) allows one to state that what is known as
the effective $1$-dimensional Lagrangian $\mathcal{L}$ for the extremal
black hole dynamics\emph{\ }given by (\ref{eq:normal_effective}) actually is
the \textit{Routhian functional} (defined by (\ref{eq:Routh_def})) of the
original Lagrangian functional $L$ occurring in \eqref{eq:theaction}. As $%
\mathbf{S}$ is the action pertaining to the full-fledged $4$-dimensional
Lagrangian $L$, and $\mathcal{S}$ is the action pertaining to the effective $%
1$-dimensional Lagrangian $\mathcal{L}$, the action pertaining to $\mathcal{R%
}$ will be denoted by $\mathfrak{S}$, and it can be defined as follows :
\begin{equation}
\mathfrak{S}:=4\pi \int_{0}^{\infty }\mathcal{R}\,d\tau =4\pi
\int_{0}^{\infty }\left( \dot{U}^{2}+{\frac{1}{2}}G_{a\bar{a}}\dot{z}^{a}%
\dot{\bar{z}}^{\bar{a}}+e^{2U}V_{BH}\right) d\tau =4\pi \mathcal{S},
\label{eq:static}
\end{equation}%
where in the last step we recalled (\ref{eq:normal_action}). Although $%
\mathcal{R}$ is a particular Routhian functional pertaining to the $4$%
-dimensional Lagrangian $L$, \textit{for all practical purposes} $\mathcal{R}
$ \textit{itself can be thought of as a mechanical Lagrangian}.

\subsubsection{The Hamiltonian constraint}

In order to derive the constraint \eqref{eq:constraint}, one should note
that the full Lagrangian $L$ in four space-time dimensions %
\eqref{eq:theaction} is generally \textit{covariant}, or in other words one
can freely choose any coordinate system to work on, and the physics would
remain the same. The effective $1$-dimensional functional $\mathcal{R}$ (\ref%
{eq:finalrouth}) does not have any explicit dependence on the `radial time' $%
\tau $; therefore, to still maintain the general covariance where one is
free to redefine $\tau $, $\mathcal{R}$ should be \textit{reparametrisation
invariant} under any redefinition $\tau \rightarrow \kappa (\tau )$. Since $%
\tau $ is a radial coordinate, the corresponding Hamiltonian functional $%
\mathcal{H}$ generates a reparametrisation, rather than time evolution :
equivalently, one may say that the physical system described by $\mathcal{R}$
does not evolve in time (like in standard mechanics), but its dynamics is
instead constrained to be governed by the Euler-Lagrange equations related
to $\mathcal{R}$, for \textit{all} allowed values of $\tau \in \mathbb{R}%
^{+} $. \textit{The only way to achieve this is to make the Hamiltonian to
be strictly zero}. Defining the momenta-like variables
\begin{equation}
\pi _{U}:={\frac{\delta \mathcal{R}}{\delta \dot{U}}}=2\dot{U};\quad \pi
_{a}:={\frac{\delta \mathcal{R}}{\delta \dot{z}^{a}}}={\frac{1}{2}}G_{a\bar{b%
}}\dot{\bar{z}}^{\bar{b}};\quad \bar{\pi}_{\bar{b}}={\frac{\delta \mathcal{R}%
}{\delta \dot{\bar{z}}^{\bar{b}}}}={\frac{1}{2}}G_{a\bar{b}}\dot{z}^{a},
\end{equation}%
the Hamiltonian can be defined by the full-fledged Legendre transform of $%
\mathcal{R}$ :
\begin{equation}
\mathcal{H}=\pi _{U}\dot{U}+\pi _{a}\dot{z}^{a}+\bar{\pi}_{\bar{a}}\dot{\bar{%
z}}^{\bar{a}}-\mathcal{R}=\dot{U}^{2}+{\frac{1}{2}}G_{a\bar{a}}\dot{z}^{a}%
\dot{\bar{z}}^{\bar{a}}-e^{2U}V_{BH}.
\end{equation}%
The condition of vanishing Hamiltonian is thus equivalent to %
\eqref{eq:constraint},
\begin{equation}
\mathcal{H}=0\Leftrightarrow \dot{U}^{2}+{\frac{1}{2}}G_{a\bar{a}}\dot{z}^{a}%
\dot{\bar{z}}^{\bar{a}}-e^{2U}V_{BH}=0,  \label{eq:H_constraint}
\end{equation}%
which is thus named \textit{Hamiltonian constraint}.\medskip

It is here worth recalling that the strict vanishing of $\mathcal{H}$ holds
for \textit{extremal} black holes (which are the ones under consideration
here), whereas for \textit{non-extremal} black holes, the Hamiltonian can be
set equal to a constant $c^2$, where $c$ is the extremality parameter\footnote{The metric for the non extremal black hole is given by \cite{Ferrara:1997tw}
\begin{eqnarray} \label{non_ext}
 ds^2 =- e^{2U(\tau)} dt^2 + e^{-2U(\tau)} \bigg[\frac{c^4}{\sinh^4 c \tau} d\tau^2 + \frac{c^2}{\sinh^2 c \tau} d\Omega_{2}^2(\theta,\phi) \bigg] \  
\end{eqnarray} where $c$ is defined as the extremality parameter with the property that $c^2=2ST$ with $S$ and $T$ being the entropy and the temperature of the blackhole. One can check that in the \emph{extremal limit} of $c\rightarrow 0$, \eqref{non_ext} boils down to the metric ansatz \eqref{eq:seed_solution} under suitable coordinate changes.}.
\begin{equation}
\mathcal{H}_{\text{non-extr.}}=c^2\in \mathbb{R}_{0}^{+}.
\end{equation}%

\section{\label{sec:Resume}\textit{R\'{e}sum\'{e}} on the Routhian formalism}

The \textit{Routhian} formalism is a hybrid variational framework that
eliminates a chosen subset of generalized velocities by a \textit{partial}
Legendre transform, while keeping the remaining degrees of freedom in
Lagrangian form. It was introduced by E. J. Routh more than $150$ years ago,
to exploit \textit{cyclic} coordinates in analytical dynamics \cite{Routh1877, Whittaker_1988}, and it is naturally interpreted today as a
concrete instance of symmetry reduction (\textquotedblleft \textit{Routh
reduction}\textquotedblright ) in geometric mechanics \cite%
{MarsdenRatiu1999,HolmSchmahStoica2009}. The Routhian formalism can be
regarded to be placed in between the Euler--Lagrange and Hamiltonian
pictures: indeed, it produces Euler--Lagrange equations for the non-cyclic
variables and Hamilton-type reconstruction equations for the cyclic ones
\cite{Arnold1989,Goldstein2001}. In other words, the Routhian formalism
reduces the order of the equations associated with cyclic coordinates by
trading them for conserved momenta.

\subsection{The \textit{partial} Legendre transform}

Let $Q$ be an $n$--dimensional configuration manifold with local coordinates
$q=(q^{1},\dots ,q^{n})$. Let $\mathbf{L}\colon TQ\rightarrow \mathbb{R}$ be
a smooth Lagrangian functional. Suppose the coordinates split as
\begin{equation*}
(q^{1},\dots ,q^{n})=(x^{1},\dots ,x^{n-k},\,y^{1},\dots ,y^{k})=(x,y),
\end{equation*}%
where the variables $y$ are \emph{cyclic}, i.e. $\mathbf{L}$ is independent
of $y$:
\begin{equation}
\frac{\delta \mathbf{L}}{\delta y^{\alpha }}(x,\dot{x},\dot{y})=0,\qquad
\alpha =1,\dots ,k.  \label{eq:cyclic}
\end{equation}%
The conjugate momenta associated with the cyclic velocities are
\begin{equation}
p_{\alpha }:=\frac{\delta \mathbf{L}}{\delta \dot{y}^{\alpha }}(x,\dot{x},%
\dot{y}).  \label{eq:pdef}
\end{equation}%
By \eqref{eq:cyclic}, the Euler--Lagrange equations for $y^{\alpha }$ reduce
to the conservation of $p_{\alpha }$ (a basic case of Noether's theorem)
\cite{Goldstein2001,Arnold1989,MarsdenRatiu1999}:
\begin{equation}
\dot{p}_{\alpha }=0\quad \Rightarrow \quad p_{\alpha }=\mu _{\alpha }\ \text{%
(constants).}  \label{eq:momentum_conservation}
\end{equation}

\subsection{Partial regularity and solvability}

In this context, one generally assumes the so-called $y$\textit{--regularity
}: namely, that the partial Hessian in the eliminated velocities,
\begin{equation}
\mathbb{H}_{\alpha \beta }:=\frac{\delta ^{2}\mathbf{L}}{\delta \dot{y}%
^{\alpha }\delta \dot{y}^{\beta }}  \label{eq:regularity}
\end{equation}%
is \textit{invertible}. This is exactly the condition ensuring that the
Legendre map in the $\dot{y}$ directions is locally a diffeomorphism, so
that \eqref{eq:pdef} can be solved for $\dot{y}$ as
\begin{equation}
\dot{y}=\dot{y}(x,\dot{x},p).  \label{eq:solve_doty}
\end{equation}%
The $y$\textit{--regularity} condition \eqref{eq:regularity} is the minimal
hypothesis ensuring that the conserved momenta can be used as coordinates to
eliminate $\dot{y}$. In this sense, it is the partial analog of the
so-called \textit{hyperregularity} in Hamiltonian mechanics \cite{Arnold1989}%
.

\subsection{Definition of the Routhian}

The \textit{Routhian} functional is the partial Legendre transform of $%
\mathbf{L}$ with respect to the cyclic velocities\footnote{%
When the Lagrangian explicitly depends on $t$ ($\mathbf{L}=\mathbf{L}(t,q,%
\dot{q})$), one may still define $\mathcal{R}(t,x,\dot{x},p)$ by the same
partial transform. In general, the conservation of $p$ persists whenever the
$y$ variables remain cyclic \cite{Goldstein2001}.} \cite%
{Routh1877,Whittaker_1988,Goldstein2001}:
\begin{equation}
\mathcal{R}(x,\dot{x},p):=\mathbf{L}\left( x,\dot{x},\dot{y}(x,\dot{x}%
,p)\right) \;-\;\sum_{\alpha =1}^{k}p_{\alpha }\,\dot{y}^{\alpha }(x,\dot{x}%
,p).  \label{eq:Rdef}
\end{equation}%
Thus, $\mathcal{R}$ is generally a smooth functional of $(x,\dot{x})$ and of
the parameters $p\in \mathbb{R}^{k}$; more rigorously, $\mathcal{R}$ is a
functional on the bundle $T(Q/G)\times \mathbb{R}^{k}$ in the simplest
(Abelian) situation.

\subsection{Mixed equations of motion and variational principle}

The key computational point is the following : since $\dot{y}$ is chosen to
satisfy the stationarity conditions
\begin{equation}
\frac{\delta }{\delta \dot{y}^{\alpha }}\left( \mathbf{L}(x,\dot{x},\dot{y}%
)-p\cdot \dot{y}\right) =0\quad \Longleftrightarrow \quad p_{\alpha }=\frac{%
\delta L}{\delta \dot{y}^{\alpha }},  \label{eq:stationarity}
\end{equation}%
the dependence $\dot{y}=\dot{y}(x,\dot{x},p)$ does \textit{not} contribute
to the first derivatives of $\mathcal{R}$ with respect to $(x,\dot{x})$.
Concretely, one finds the identities (see e.g. \cite%
{Arnold1989,Goldstein2001}), $i=1,...,n-k$ :
\begin{align}
\frac{\delta \mathcal{R}}{\delta \dot{x}^{i}}& =\frac{\delta \mathbf{L}}{%
\delta \dot{x}^{i}}\Big|_{\dot{y}=\dot{y}(x,\dot{x},p)};  \label{eq:dRdxdot}
\\
\frac{\delta \mathcal{R}}{\delta x^{i}}& =\frac{\delta \mathbf{L}}{\delta
x^{i}}\Big|_{\dot{y}=\dot{y}(x,\dot{x},p)}.  \label{eq:dRdx}
\end{align}

\subsection{Reduced Euler--Lagrange equations}

The \textit{reduced} Lagrangian functional $\mathbf{L}_{\mu }$ can be
defined as the Routhian functional $\mathcal{R}$ with the momentum
conservation condition (\ref{eq:momentum_conservation}) implemented :
\begin{equation}
\mathbf{L}_{\mu }(x,\dot{x}):=\left. \mathcal{R}(x,\dot{x},p)\right\vert _{%
\text{(\ref{eq:momentum_conservation})}}=\mathcal{R}(x,\dot{x},\mu ).
\label{eq:Lmu}
\end{equation}%
Then, the $x$--components of the Euler--Lagrange equations for $\mathbf{L}$
are equivalent to the Euler--Lagrange equations for $\mathbf{L}_{\mu }$ \cite%
{Routh1877,Whittaker_1988,MarsdenRatiu1999}:
\begin{equation}
\frac{d}{dt}\frac{\delta \mathbf{L}_{\mu }}{\delta \dot{x}^{i}}-\frac{\delta
\mathbf{L}_{\mu }}{\delta x^{i}}=0,~~i=1,\dots ,n-k.  \label{eq:EL_reduced}
\end{equation}

\subsection{Reconstruction (Hamilton-type) equations}

The eliminated variables satisfy first-order relations, obtained by
differentiating $\mathcal{R}$ with respect to $p$:
\begin{equation}
\dot{y}^{\alpha }=-\frac{\partial \mathcal{R}}{\partial p_{\alpha }}(x,\dot{x%
},p).  \label{eq:ydot_from_R}
\end{equation}%
These are the analogues of Hamilton's equations $\dot{q}=\delta H/\delta p$
for the partially transformed directions; in fact, if $k=n$ then the
Routhian $\mathcal{R}$ becomes the negative of the full Hamiltonian
formalism (up to the sign convention in \eqref{eq:Rdef}) \cite{Arnold1989}.
Moreover, (\ref{eq:ydot_from_R}) become explicit \textit{reconstruction
equations} once $x(t)$ is known. An example of application of the Routhian
formalism (the so-called `planar central-force problem' is given in App. \ref%
{App-Example-Routh}), whereas further subtleties concerning `magnetic' terms
possibly appearing in the reduced, effective Routhian functional are briefly
discussed in App. \ref{App-Magn}.

\section{\label{sec:Attractors-Entropy}Attractors and entropy}

\subsection{Attractors as critical points of $V_{BH}$}

Let us now consider the equations of motion of the scalar fields $z^{a}$,
whose radial dynamics (i.e., the only existing dynamics!) is governed by the
Routhian $\mathcal{R}$ given by \eqref{eq:finalrouth} : the corresponding
Euler-Lagrange (i.e., variational) equations (usually named equations of
motion) for $\bar{z}^{\bar{a}}$ read
\begin{equation}
\frac{d}{d\tau }\left( \frac{\delta \mathcal{R}}{\delta \left( \partial
_{\tau }\bar{z}^{\bar{a}}\right) }\right) -\frac{\delta \mathcal{R}}{\delta
\bar{z}^{\bar{a}}}=0\ ,  \label{eq:scalar_eom1}
\end{equation}%
or explicitly,
\begin{equation}
\frac{1}{2}\left[ G_{a\bar{b}}\partial _{\tau }^{2}z^{a}+\left( \frac{%
\partial G_{a\bar{b}}}{\partial \bar{z}^{\bar{a}}}-\frac{\partial G_{a\bar{a}%
}}{\partial \bar{z}^{\bar{b}}}\right) \partial _{\tau }z^{a}\partial _{\tau }%
\bar{z}^{\bar{a}}+\frac{\partial G_{a\bar{b}}}{\partial z^{b}}\partial
_{\tau }z^{a}\partial _{\tau }z^{b}\right] -e^{2U}\frac{\partial V_{BH}}{%
\partial \bar{z}^{\bar{b}}}=0\ .  \label{eq:scalar_eom2}
\end{equation}%
As briefly mentioned above, in the present treatment we assume the scalar
manifold (i.e., the target space of the scalar fields) to be a \textit{%
complex} Riemannian space, and more in particular a \textit{K\"{a}hler}
space : thus, its metric $G_{a\bar{b}}$ is given in terms of a real K\"{a}%
hler potential $K$ as
\begin{equation}
G_{a\bar{b}}=\frac{\partial ^{2}K}{\partial z^{a}\partial \bar{z}^{\bar{b}}}%
\ .  \label{eq:scalar_eom3}
\end{equation}%
Plugging \eqref{eq:scalar_eom3} into \eqref{eq:scalar_eom2}, and making some
simplifications, the equations of motion of of the scalar fields $z^{a}$
read
\begin{equation}
\partial _{\tau }^{2}z^{a}+\Gamma _{bc}^{a}\partial _{\tau }z^{b}\partial
_{\tau }z^{c}-e^{2U}G^{a\bar{b}}\frac{\partial V_{BH}}{\partial \bar{z}^{%
\bar{b}}}=0\ ,  \label{eom4}
\end{equation}%
where the Christoffel symbols are given by
\begin{equation}
\Gamma _{bc}^{a}=G^{a\bar{d}}\partial _{\bar{z}^{\bar{d}}}\partial
_{z^{b}}\partial _{z^{c}}K\ .
\end{equation}%
After some rearranging, Eq. \eqref{eom4} can be recast in the following form
:
\begin{equation}
\partial _{\tau }^{2}z^{a}+\Gamma _{bc}^{a}\partial _{\tau }z^{b}\partial
_{\tau }z^{c}=e^{2U}G^{a\bar{b}}\frac{\partial V_{BH}}{\partial \bar{z}^{%
\bar{b}}}\ .  \label{eom6}
\end{equation}%
This is the equation of motion of the complex scalar fields $z^{a}$ (i.e. of
the holomorphic scalar maps coordinatizing the K\"{a}hler manifold under
consideration).

In the following treatment, we will see that the left hand side of (\ref%
{eom6}) vanishes identically at the unique black hole event horizon, which
resides at $\tau \rightarrow \infty $ (whereas the asymptotical limit
corresponds to $\tau \rightarrow 0^{+}$). We should here also recall that,
in order to get a finite horizon area (and thus, through the
Bekenstein-Hawking formula, a finite entropy) from the seed solution %
\eqref{eq:seed_solution}, one should have the following near-horizon
behaviour of the warp factor,
\begin{equation}
\left. e^{-2U}\right\vert _{\tau \rightarrow \infty }={\frac{A}{4\pi }}\tau
^{2},  \label{eq:at_horizon}
\end{equation}%
where $A$ is defined as the area of the event horizon. By changing the
radial coordinate as%
\begin{equation}
\tau \longrightarrow \omega :=\log \tau \Rightarrow \frac{\partial }{%
\partial \tau }=\frac{\partial }{\partial \omega }\frac{\partial \omega }{%
\partial \tau }\frac{1}{\tau }\frac{\partial }{\partial \omega },
\end{equation}%
Eq. \eqref{eom6} boils down to
\begin{equation}
\partial _{\omega }^{2}z^{a}+\Gamma _{bc}^{a}\partial _{\omega
}z^{b}\partial _{\omega }z^{c}=G^{a\bar{b}}\frac{\partial V_{BH}}{\partial
\bar{z}^{\bar{b}}}\ .  \label{eq:eom7}
\end{equation}%
The radial coordinate $\omega $ can be considered as the \textit{physical
distance} from the event horizon in units of the event horizon radius $r_{H}$%
, since, at any given time, the interval $ds^{2}=r_{H}^{2}\,d\omega ^{2}$.
Therefore, one can of course expect that the scalar fields $z^{a}$ and all
their derivatives $\partial _{\omega }z^{a}$ should tend to some \textit{%
finite} value at the horizon, i.e. $\omega \rightarrow \infty $. In the
case, in which $\partial _{\omega }z^{a}$ takes some non-zero value at $%
\omega \rightarrow \infty $, that would inevitably lead to $z^{a}\rightarrow
\infty $, as one approaches the horizon, thus implying the scalar field to
diverge. The unique possibility to avoid such a divergence of scalar fields
at the black hole event horizon is setting, $\forall a=1,...,n$,
\begin{eqnarray}
\text{lim}_{\tau \rightarrow \infty }z^{a}\left( \tau \right)
&=&:z_{H}^{a},~\left\vert z_{H}^{a}\right\vert <\infty ~;  \label{this} \\
\text{lim}_{\tau \rightarrow \infty }\dot{z}^{a}\left( \tau \right) &=&0.
\label{this-2}
\end{eqnarray}%
An analogous argument by induction yields to conclude that all higher-order
derivatives of the scalar fields $z^{a}$ should also vanish at the horizon.
The constant values $z_{H}^{a}$ of the scalar fields at the horizon is a
critical value for $V_{BH}$ itself :
\begin{equation}
\frac{\partial V_{BH}}{\partial \bar{z}^{\bar{b}}}|_{\left\{
z^{a}=z_{H}^{a}\right\} _{a=1,...,n}}=0,~\forall b=1,...,n.
\label{crit_cond}
\end{equation}%
It is straightforward to observe that the values $z_{H}^{a}$ defined by (\ref%
{this}) which solve the criticality condition (\ref{crit_cond}) for $V_{BH}$
is \textit{independent} of its initial conditions of the dynamical system
describing its radial evolution, from spacial infinity ($\tau \rightarrow
0^{+}$) to the event horizon ($\tau \rightarrow \infty $), namely from the
asymptotical values of $z^{a}$ and\footnote{%
These latter are sometimes called (asymptotical) \textit{scalar charges}.} $%
\dot{z}^{a}$. Thus, borrowing the terminology from the mathematical theory
of dynamical systems, one obtains that the black hole event horizon (which
is unique due to the extremality of the black hole itself) behaves as an
\textit{attractor point}\footnote{%
When the scalar manifold is not simply connected (thus, non-homogeneous),
some `area codes' or `basins of attraction' may arise. Thus, the attractor
mechanism gets spoiled globally, but it holds only \textit{at most} locally; see e.g. \cite{Kallosh:1999mz,Giryavets:2005nf}.} for the $n$ ordinary differential equations %
\eqref{eq:eom7}; the `attracted' values of the scalar fields at the horizon,
denoted as $z_{H}^{a}$, depend only on the $2n+2$ real conserved charges,
which are the aforementioned \textit{magnetic-electric black hole charges},
preserved by the $U(1)^{n+1}$gauge symmetry of the system, and arranged into
the symplectic vector $\mathcal{Q}$.

\subsection{\label{subsec:BH-W}$S_{BH}=S_{W}$ for two-derivative gravity}

To compute the entropy of the extremal black hole seed solution %
\eqref{eq:seed_solution}, one needs to focus on the near horizon ($\tau
\rightarrow \infty $) region. As discussed above, the horizon is an \textit{%
attractor} for the scalar fields (which acquire purely charge-dependent,
finite values at the horizon), and all of their derivatives also vanish
there (cf. (\ref{this-2})). Thus, by recalling the Bekenstein-Hawking
entropy-area formula%
\begin{equation}
S_{BH}=\frac{A}{4}  \label{S_BH}
\end{equation}%
(with $S_{BH}$ denoting the entropy of the extremal black hole), the
Hamiltonian constraint \eqref{eq:H_constraint} and the observation %
\eqref{eq:at_horizon} yield, at the horizon, the following relation between
the Bekenstein-Hawking entropy $S_{BH}$ and the horizon value $\left.
V_{BH}\right\vert _{\text{hor}}$ of the effective black hole potential $%
V_{BH}$ (\ref{vbh}) :
\begin{equation}
S_{BH}=\frac{A}{4}=\pi \left. V_{BH}\right\vert _{\text{hor}}\ =\pi \left.
V_{BH}\right\vert _{\partial _{z^{a}}V_{BH}=0~\forall a},
\label{eq:entropy_potential}
\end{equation}%
where in the last step we exploited the criticality conditions for $V_{BH}$,
expressed by (\ref{crit_cond}). Hence, the (Bekenstein-Hawking) entropy of
the extremal black hole solution \eqref{eq:seed_solution} is determined by
the critical points of the functional $V_{BH}$.

Another approach to determine the entropy\ is based on the Wald formula :
within such a prescription, the black hole entropy $S_{W}$ is given by \cite%
{Iyer:1994ys}
\begin{equation}
S_{W}:={\frac{1}{4}}\int_{H}d^{2}x\,\sqrt{h}{\frac{\delta L}{\delta R_{\mu
\nu \rho \sigma }}}\epsilon _{\mu \nu }\epsilon _{\rho \sigma },
\label{eq:wald_def}
\end{equation}%
where $H$ denotes the (2-dimensional) horizon of the (4-dimensional) black
hole, $h$ being the 2-dimensional induced metric on the horizon, and $%
\epsilon _{\mu \nu }$ denoting the bi-normal to the horizon with the
normalisation $\epsilon _{\mu \nu }\epsilon ^{\mu \nu }=-2$. Since for the
computation of the functional derivative in (\ref{eq:wald_def}) the relevant
part of the Lagrangian $L$ (cf. (\ref{eq:theaction})) is only the
Einstein-Hilbert one ($-R/2$), one can compute
\begin{equation}
{\frac{\delta L}{\delta R_{\mu \nu \rho \sigma }}}=-{\frac{1}{4}}\left(
g^{\mu \rho }g^{\nu \sigma }-g^{\mu \sigma }g^{\nu \rho }\right) .
\end{equation}%
Therefore, one obtains
\begin{equation}
S_{W}=S_{BH}={\frac{A}{4}}=\pi \,\left. V_{BH}\right\vert _{\text{hor}},
\label{Rel1}
\end{equation}%
a result consistently equal to the Bekenstein-Hawking formula for the
extremal black hole, which in (\ref{eq:entropy_potential}) has been
implemented in terms of the critical values of $V_{BH}$. Thus, the formula (%
\ref{Rel1}) relates the critical values of the black hole effective
potential $V_{BH}$ to the Bekenstein-Hawking entropy $S_{BH}$, which in
turn, in the framework under consideration, matches the black hole entropy $%
S_{W}$ computed by applying Wald's formula.

Next, we are going to analyze the relation of all these with Sen's entropy
function(al) formalism.

\section{A \textit{trio} of functionals : $\mathcal{E}$, $\mathcal{R}$, and $%
V_{BH}$}\label{sec:efd}

 To understand the relation between Sen's entropy functional,
introduced in \cite{Sen:2005wa}, and the formalism based on the black hole
effective potential $V_{BH}$, dating back to \cite{Ferrara:1997tw}, it is
instructive to scrutinize Sen's prescription \cite{Sen:2005wa} for the
computation of the entropy from the Maxwell-Einstein-scalar action %
\eqref{eq:theaction}, pertaining to the asymptotically flat, static,
spherically symmetric, extremal black hole solution given by %
\eqref{eq:seed_solution}.

\subsection{\label{subsec:BR}Bertotti-Robinson $AdS_{2}\times S^{2}$
Near-horizon geometry}

Sen's entropy functional formalism is inherently near-horizon, namely it
focuses on the near-horizon limit of the extremal black hole metric %
\eqref{eq:seed_solution}. By exploiting the property expressed by %
\eqref{eq:at_horizon} and performing the coordinate change $\tau \rightarrow
{\frac{1}{r}}$, one obtains
\begin{eqnarray}
ds^{2} &=&{\frac{1}{\mathbf{v}}}r^{2}dt^{2}-\mathbf{v}\left[ {\frac{dr^{2}}{%
r^{2}}}+\left( d\theta ^{2}+\sin ^{2}\theta \,d\phi ^{2}\right) \right] ;
\label{eq:nearH} \\
\mathbf{v} &:=&{\frac{A}{4\pi }}={\frac{S_{BH}}{\pi }=}\left.
V_{BH}\right\vert _{\text{hor}}.  \label{eq:nearH2}
\end{eqnarray}%
This is the near-horizon limit of the metric \eqref{eq:seed_solution}, and
it goes under the name of Bertotti-Robinson $AdS_{2}\times S^{2}$ geometry,
a conformally flat factorized metric with spherical 2-dimensional horizon,
and \cite{Bertotti1959,Robinson1959}
\begin{equation}
r_{AdS_{2}}^{2}=r_{S^{2}}^{2}=\mathbf{v}.
\end{equation}%
In \cite{Sen:2005wa}, Sen postulated that \textit{in any generally covariant
theory of gravity coupled to matter fields, the near-horizon geometry of an
asymptotically flat, spherically symmetric, static extremal black hole in
four dimensions has}\emph{\ }$SO(2,1)\times SO(3)$\emph{\ }\textit{isometry}
(namely, it is of Bertotti-Robinson type\footnote{%
The $AdS_{2}$ part (with $SO(2,1)$ isometry) is preserved also for
asymptotically non-flat \cite{Kunduri:2013gce}, \textit{and/or} non-static
(but stationary), extremal BHs \cite{Kunduri:2008tk} }); to make the $%
SO(2,1)\times SO(3)$ global symmetry of the near-horizon geometry %
\eqref{eq:nearH} manifest, it suffices to rescale the time coordinate as
\begin{equation}
t\rightarrow {\frac{A}{4\pi }}\hat{t}=\mathbf{v}\hat{t}\Rightarrow ds^{2}=%
\mathbf{v}\left[ r^{2}d\hat{t}^{2}-{\frac{dr^{2}}{r^{2}}}-\left( d\theta
^{2}+\sin ^{2}\theta \,d\phi ^{2}\right) \right] ,
\end{equation}%
or, equivalently, to rescale the radial coordinate as%
\begin{equation}
r\rightarrow \mathbf{v}\hat{r}\Rightarrow ds^{2}=\mathbf{v}\left[ \hat{r}%
^{2}dt^{2}-{\frac{d\hat{r}^{2}}{\hat{r}^{2}}}-\left( d\theta ^{2}+\sin
^{2}\theta \,d\phi ^{2}\right) \right] .
\end{equation}

For completeness, a thorough treatment (originally given by Sen in \cite%
{Sen:2005wa}) dealing with the most general field configuration consistent
with the $SO(2,1)\times SO(3)$ isometry (which actually is slight more
general of the above one, for what concerns the gravity sector) for the
action \eqref{eq:theaction} is reviewed in \cref{app:genRN}.

\subsection{\label{f}The functional $f$}

After \cite{Sen:2005wa}, one can infer that in the near-horizon region,
described by (\ref{eq:nearH})-(\ref{eq:nearH2}), the scalar fields $z^{a}$
take constant values, whereas the unique non-vanishing components of the
Abelian 2-form field strengths are given by{\footnote{%
It is worthwhile to mention that here the coordinate system is different
from the one adopted in the previous section. However, by applying the
transformation rules for rank-2 antisymmetric tensors (i.e., 2-forms) under
general coordinate transformations, one can show that at the horizon the
components $\left. F_{rt}^{\Lambda }\right\vert _{\text{hor}}$ as well as $%
\left. F_{\theta \phi }^{\Lambda }\right\vert _{\text{hor}}$ acquire the
same values they had in the treatment of the previous section.}}%
\begin{eqnarray}
F_{rt}^{\Lambda } &=&e^{\Lambda };  \label{F1} \\
F_{\theta \phi }^{\Lambda } &=&-p^{\Lambda }\sin \theta .  \label{F2}
\end{eqnarray}

Then, one can define the real functional $f(z^{a},\bar{z}^{\bar{a}%
};e^{\Lambda },p^{\Lambda })$ as the integration of the Lagrangian density $L
$ (evaluated on the near-horizon metric\footnote{%
Contrary to the (slightly more general) treatment given in \cref{app:genRN},
here the functional $f$ is inherently defined over the Bertotti-Robinson
near-horizon metric \eqref{eq:nearH} (in which the radius of the $AdS_{2}$
factor coincides with the radius of the $2$-sphere factor $S^{2}$).} (\ref%
{eq:nearH})-(\ref{eq:nearH2})) over the angular coordinates ($\theta $ and $%
\phi $) :
\begin{eqnarray}
f(z^{a},\bar{z}^{\bar{a}};e^{\Lambda },p^{\Lambda }) &:&=\int d\theta d\phi
\ \sqrt{-g}\left. L\right\vert _{\text{(\ref{eq:nearH})-(\ref{eq:nearH2})}}
\notag \\
&=&-2\pi \,\mathbf{v}\,\mu _{\Lambda \Sigma }\,\left( z^{a},\bar{z}^{\bar{a}%
}\right) e^{\Lambda }e^{\Sigma }+\frac{2\pi }{\mathbf{v}}\mu _{\Lambda
\Sigma }\,\left( z^{a},\bar{z}^{\bar{a}}\right) p^{\Lambda }p^{\Sigma }-4\pi
\,\nu _{\Lambda \Sigma }\,\left( z^{a},\bar{z}^{\bar{a}}\right) e^{\Lambda
}p^{\Sigma }\ ,\notag 
\\ 
\label{eq:f_def}
\end{eqnarray}%
where evaluation in the near-horizon region (\ref{eq:nearH})-(\ref{eq:nearH2}%
) is understood in the second line.

The extremization of $f$ with respect to the scalar fields yields the corresponding equations of motion 
\begin{equation}
{\frac{\delta f}{\delta z^{a}}}=0\Leftrightarrow \,\frac{\partial \mu
_{\Lambda \Sigma }}{\partial z^{a}}\,\left( -\,\mathbf{v}e^{\Lambda
}e^{\Sigma }+\frac{1}{\mathbf{v}}p^{\Lambda }p^{\Sigma }\right) \,-2\,\frac{%
\partial \nu _{\Lambda \Sigma }}{\partial z^{a}}\,e^{\Lambda }p^{\Sigma }=0.
\label{eq:crit-f-z}
\end{equation}%
On the other hand, although the gauge fields' equations of motion as well as
the Bianchi identities\footnote{%
From the gauge equations of motion ${\frac{1}{\sqrt{-g}}}\partial _{\mu
}\left( \sqrt{-g}\,\mathcal{G}_{\Lambda }^{\mu \nu }\right) =0\implies
\partial _{r}\left( {\frac{\delta L}{\delta F_{rt}^{\Lambda }}}\right) =0$
and the Bianchi identity implies $\partial _{r}F_{\theta \phi }^{\Lambda
}=0. $} are automatically satisfied by the background solutions (\ref{F1})-(%
\ref{F2}) for the Abelian 2-form field strengths, one can use the full
Lagrangian $L$ (whose integration over space-time yields the action $\mathbf{%
S}$) in order to define the black hole electric charge $q_{\Lambda }$. In
fact, from the relation in \eqref{eq:field_pot}, the electric charge of the
black hole, defined as the flux of the dual 2-form $\mathcal{G}_{\Lambda }$
through the spherical horizon $S^{2}$ of the Bertotti-Robinson near-horizon
geometry (\ref{eq:nearH}),
\begin{equation}
q_{\Lambda }:=-\frac{1}{4\pi }\int_{S^{2}}\mathcal{G}_{\Lambda \theta \phi },
\end{equation}%
can also be written as
\begin{equation}
q_{\Lambda }=\frac{1}{4\pi }\frac{\partial f}{\partial e^{\Lambda }}=-\left(
\mathbf{v}\,\mu _{\Lambda \Sigma }e^{\Sigma }+\nu _{\Lambda \Sigma
}p^{\Sigma }\right) \ ,  \label{eq:qlambda_def}
\end{equation}%
where we used the definition of $\mathcal{G}_{\Lambda }$ given by (\ref%
{eq:field_pot_2}) as well as the relation \eqref{eq:f_def}. In other words,
the black hole electric charge $q_{\Lambda }$ is ($1/4\pi $ times) the
momentum-like coordinate \textit{conjugate} to $e^{\Lambda }$; this latter
can thus be regarded as an `auxiliary' magnetic variable (actually,
instrumental to introduce $q_{\Lambda }$ by dualization or, equivalently,
canonical conjugation).

It is here worth remarking that, contrary to the usual lore, the equations \eqref{eq:crit-f-z}, obtained  by extremizing $f$, are \emph{not} equivalent to the correct scalar equations of motion \eqref{eq:eoms2} (in the near horizon limit in which the scalar fields take constant values), which boil down to \eqref{crit_cond}, thus implying
\begin{equation}
\frac{\partial V_{BH}}{\partial \bar{z}^{\bar{b}}}=0\Leftrightarrow \,\frac{\partial \mu
_{\Lambda \Sigma }}{\partial z^{a}}\,\left( -\,\mathbf{v}e^{\Lambda
}e^{\Sigma }+\frac{1}{\mathbf{v}}p^{\Lambda }p^{\Sigma }\right)=0,
\end{equation}
where the definition \eqref{eq:qlambda_def} has been used. Hence, the equations of motion of the scalar fields match the equations \eqref{eq:crit-f-z} if and only if%
\begin{equation}
\frac{\partial \nu _{\Lambda \Sigma }}{\partial z^{a}}e^{\Lambda }p^{\Sigma
}=0,
\end{equation}%
thus, for instance, when $\nu _{\Lambda \Sigma }=0$, namely in absence of the would-be topological term in the four-dimensional starting Lagrangian, as it is indeed the case for \cite{Sen:2007qy,Sen:2005wa}. This mismatch concerning the scalar equations of motion, although not seen while using a simpler Lagrangian without the topological term, is easily resolved by using Sen's entropy functional $\mathcal{E}$ approach, as discussed in the next section.

\subsection{\label{E}Sen's entropy functional $\mathcal{E}$}

Sen's entropy functional $\mathcal{E}$ is defined as the
Legendre transform of the functional $f$ (\ref{eq:f_def}) with respect to $%
e^{\Lambda }$ :
\begin{equation}
\mathcal{E}(z^{a},\bar{z}^{\bar{a}};e^{\Lambda },p^{\Lambda },q_{\Lambda }):=%
\mathfrak{n}\left( 4\pi e^{\Lambda }q_{\Lambda }-f(z^{a},\bar{z}^{\bar{a}%
};e^{\Lambda },p^{\Lambda })\right) ,  \label{eq:ent_def}
\end{equation}%
where $\mathfrak{n}$ is a (crucial!) normalization constant. As all the
treatment of this Section, and as the functional $f$, also the functional $%
\mathcal{E}$ is \textit{inherently defined in the near-horizon region}
described by (\ref{eq:nearH})-(\ref{eq:nearH2}).

Both the scalar equations of motion and the definition \eqref{eq:qlambda_def}
of the electric charge $q_{\Lambda }$ (as conjugate to the `auxiliary'
magnetic variable $e^{\Lambda }$) can be derived by extremizing $\mathcal{E}$
with respect to $z^{a}$ and $e^{\Lambda }$, respectively :%
\begin{eqnarray}
{\frac{\delta \mathcal{E}}{\delta z^{a}}} &=&%
0;  \label{critt} \\
{\frac{\partial \mathcal{E}}{\partial e^{\Lambda }}} &=&0\Leftrightarrow
q_{\Lambda }=\frac{1}{4\pi }\frac{\partial f}{\partial e^{\Lambda }}.
\label{critt2}
\end{eqnarray}

As a crucial feature of the entropy function(al) formalism, in \cite%
{Sen:2005wa} Sen chooses to fix $\mathfrak{n}$ by requiring that, at its
critical points (\ref{critt})-(\ref{critt2}), $\mathcal{E}$ yields the
extremal black hole entropy :%
\begin{equation}
\left. \mathcal{E}\right\vert _{\text{(\ref{critt})-(\ref{critt2})}}=%
\mathfrak{n}\left( e^{\Lambda }\frac{\partial f}{\partial e^{\Lambda }}%
-f\right) _{\text{(\ref{critt})-(\ref{critt2})}}=S_{W}=S_{BH}.  \label{This}
\end{equation}%
Thus, up to a (crucial! see below) normalization constant $\mathfrak{n}$,
the Wald entropy or, equivalently, the Bekenstein-Hawking entropy, should be
regarded as the Legendre transform of the functional $f$ (\ref{eq:f_def})
with respect to $e^{\Lambda }$, evaluated at the criticality conditions (\ref%
{critt})-(\ref{critt2}) of Sen's entropy functional $\mathcal{E}$ itself.

In order to prove (\ref{This}), let us continue to follow\footnote{%
Note that the convention/factors are appropriately different in our
treatment.} \cite{Sen:2005wa}, and let us focus on the Wald prescription %
\eqref{eq:wald_def} to compute the black hole entropy,
\begin{equation}
S_{W}:=\int d\theta \,d\phi {\frac{\delta L}{\delta R_{rtrt}}}\sqrt{%
-g_{rr}g_{tt}}.=-g_{rr}g_{tt}{\frac{\delta L}{\delta R_{rtrt}}}A,
\label{eq:wald_2}
\end{equation}%
where the background metric is the Bertotti-Robinson one, given by (\ref%
{eq:nearH})-(\ref{eq:nearH2}). Let us now define the auxiliary, $\lambda $%
-parameterized functional
\begin{equation}
f_{\lambda }:=\int d\theta d\phi \ \sqrt{-g}\left. L_{\lambda }\right\vert _{%
\text{(\ref{eq:nearH})-(\ref{eq:nearH2})}},
\end{equation}%
where $L_{\lambda }$ is the deformed version of the Lagrangian $L$, in which
any factor of $R_{\alpha \beta \gamma \delta }$ is scaled by a factor $%
\lambda \in \mathbb{R}_{0}^{+}$. One then obtains that
\begin{equation}
\left. {\frac{\partial f_{\lambda }}{\partial \lambda }}\right\vert
_{\lambda =1}=-\int d\theta \,d\phi \sqrt{-g}R_{\alpha \beta \gamma \delta }{%
\frac{\delta L}{\delta R_{\alpha \beta \gamma \delta }}}.  \label{eq:fl_der}
\end{equation}%
By a little algebra (prescribed and detailed in \cite{Sen:2005wa}; see also %
\cref{app:genRN}), one can observe that ${\delta }L{/\delta R_{\alpha \beta
\gamma \delta }}$ is proportional to $g^{\alpha \gamma }g^{\beta \delta
}-g^{\alpha \delta }g^{\beta \gamma }$, with the proportionality constant
easily determined to read $-{\delta }L{/\delta R_{rtrt}}$, thus yielding to
\begin{equation}
{\frac{\delta L}{\delta R_{\alpha \beta \gamma \delta }}}=-(g^{\alpha \gamma
}g^{\beta \delta }-g^{\alpha \delta }g^{\beta \gamma }){\frac{\delta L}{%
\delta R_{rtrt}}}.
\end{equation}%
Using this relation back into \eqref{eq:fl_der}, one obtains
\begin{equation}
\left. {\frac{\partial f_{\lambda }}{\partial \lambda }}\right\vert
_{\lambda =1}=-{\frac{4}{\mathbf{v}}}{\frac{\delta L}{\delta R_{rtrt}}}\,A,
\end{equation}%
which, plugged back into \eqref{eq:wald_2}, yields to a quite elegant
expression of the Wald entropy in this framework :
\begin{equation}
S_{W}=-{\frac{\mathbf{v}}{4}}\left. {\frac{\partial f_{\lambda }}{\partial
\lambda }}\right\vert _{\lambda =1}.  \label{eq:wald_Routhain}
\end{equation}%
This expression relates the Wald entropy to\footnote{%
Note that the r.h.s. of (\ref{eq:wald_Routhain}) does not depend on scalar
fields, since we work in the Einstein frame (in which the gravitational part
of the Lagrangian is the Einstein-Hilbert Lagrangian density $-R/2$, cf. (%
\ref{eq:theaction})), in which there is minimal coupling between the scalar
fields and the space-time curvature $R$.} a certain \textit{deformation} of
the functional $f$, whose Legendre transform (with respect to $e^{\Lambda }$%
) defines Sen's entropy function(al) $\mathcal{E}$.

In order to relate (\ref{eq:wald_Routhain}) with (\ref{This}), we consider
the following reasoning \cite{Sen:2005wa} : in order to ensure invariance
under general changes of the non-angular coordinates $r$ and $t$, the
functional $f_{\lambda }$ should obey the following Euler-like, $\lambda $%
-parametrized equation :
\begin{equation}
\lambda {\frac{\partial f_{\lambda }}{\partial \lambda }}+e^{\Lambda }{\frac{%
\partial f_{\lambda }}{\partial e^{\Lambda }}}=f_{\lambda },
\end{equation}%
i.e., the functional $f_{\lambda }$ should be homogeneous of degree $1$ in $%
\lambda $ and $e^{\Lambda }$. By setting $\lambda =1$, this implies
\begin{equation}
\left. \lambda {\frac{\partial f_{\lambda }}{\partial \lambda }}\right\vert
_{\lambda =1}=\left. {\frac{\partial f_{\lambda }}{\partial \lambda }}%
\right\vert _{\lambda =1}=-\left( e^{\Lambda }{\frac{\partial f_{\lambda }}{%
\partial e^{\Lambda }}}-f_{\lambda }\right) _{\lambda =1}=-e^{\Lambda
}\left. {\frac{\partial f_{\lambda }}{\partial e^{\Lambda }}}\right\vert
_{\lambda =1}+f,  \label{eq:Sen_Routhian}
\end{equation}%
where we used $f_{\lambda =1}=f$. By further observing that (cf. the
footnote just below (\ref{eq:wald_Routhain}))%
\begin{equation}
\left. {\frac{\partial f_{\lambda }}{\partial e^{\Lambda }}}\right\vert
_{\lambda =1}=\frac{\partial f}{\partial e^{\Lambda }},
\end{equation}%
(\ref{eq:Sen_Routhian}) can be rewritten as\footnote{%
Since the l.h.s. of (\ref{eq:Sen_Routhian_2}) is independent on the scalar
fields (cf. the footnote just below (\ref{eq:wald_Routhain})), its r.h.s. is
necessarily evaluated at the criticality conditions ${\frac{\delta f}{\delta
z^{a}}}=0$ (cf. (\ref{eq:crit-f-z})). Moreover, due to (\ref%
{eq:wald_Routhain}) and to the fact that the entropy of a dyonic black hole
(as it is the case under consideration) does depend on both its magnetic and
electric charges, also the criticality condition (\ref{critt2}) of $\mathcal{%
E}$ with respect to $e^{\Lambda }$ (introducing the electric charge $%
q_{\Lambda }$) should be implemented on the r.h.s. of (\ref%
{eq:Sen_Routhian_2}).}%
\begin{equation}
\left. {\frac{\partial f_{\lambda }}{\partial \lambda }}\right\vert
_{\lambda =1}=-e^{\Lambda }\frac{\partial f}{\partial e^{\Lambda }}+f,
\label{eq:Sen_Routhian_2}
\end{equation}%
whose r.h.s. is the opposite of the Legendre transform of $f$ with respect
to $e^{\Lambda }$. Consequently, the Wald entropy $S_{W}$ (\ref%
{eq:wald_Routhain}) can be rewritten as :
\begin{equation}
S_{W}=-{\frac{\mathbf{v}}{4}}\left. {\frac{\partial f_{\lambda }}{\partial
\lambda }}\right\vert _{\lambda =1}={\frac{\mathbf{v}}{4}}\left( e^{\Lambda }%
{\frac{\partial f}{\partial e^{\Lambda }}}-f\right) _{\text{(\ref{critt})-(%
\ref{critt2})}}=\frac{\mathbf{v}}{4\mathfrak{n}}\left. \mathcal{E}%
\right\vert _{\text{(\ref{critt})-(\ref{critt2})}}.
\label{eq:entropy_legendre}
\end{equation}%
Therefore, in order for one of the equalities of (\ref{This}), namely $%
\left. \mathcal{E}\right\vert _{\text{(\ref{critt})-(\ref{critt2})}}=S_{W}$,
to hold true, it must hold that%
\begin{equation}
\left. \mathcal{E}\right\vert _{\text{(\ref{critt})-(\ref{critt2})}%
}=S_{W}\Leftrightarrow \frac{\mathbf{v}}{4\mathfrak{n}}=1\Leftrightarrow
\mathfrak{n}=\frac{\mathbf{v}}{4}\overset{\text{(\ref{eq:nearH2})}}{=}\frac{%
S_{BH}}{4\pi }\overset{\text{(\ref{S_BH})}}{=}{\frac{A}{16\pi }}\overset{%
\text{(\ref{Rel1})}}{{=}}\frac{V_{BH,\text{hor}}}{4},  \label{Rel2}
\end{equation}%
which fixes the normalization constant $\mathfrak{n}$.

Some observations are in order :

\begin{enumerate}
\item The result (\ref{Rel2}) yields the following definition of Sen's
entropy functional (\ref{eq:ent_def}) :%
\begin{equation}
\mathcal{E}:=\frac{\mathbf{v}}{4}\left( 4\pi e^{\Lambda }q_{\Lambda
}-f\right) =\frac{S_{BH}}{4\pi }\left( 4\pi e^{\Lambda }q_{\Lambda
}-f\right) ,  \label{eq:ent_def2}
\end{equation}%
which holds in the near-horizon region of the extremal black hole seed
solution (\ref{eq:seed_solution}), described by the Bertotti-Robinson metric
(\ref{eq:nearH})-(\ref{eq:nearH2}).

\item Then, at the critical points of $\mathcal{E}$, defined by (\ref{critt}%
)-(\ref{critt2}), the formula (\ref{eq:ent_def2}) becomes
\begin{eqnarray}
\left. \mathcal{E}\right\vert _{\text{(\ref{critt})-(\ref{critt2})}} &=&%
\frac{\mathbf{v}}{4}\left( 4\pi e^{\Lambda }q_{\Lambda }-f\right) _{\text{(%
\ref{critt})-(\ref{critt2})}}={\frac{\mathbf{v}}{4}}\left( e^{\Lambda }{%
\frac{\partial f}{\partial e^{\Lambda }}}-f\right) _{\text{(\ref{critt})-(%
\ref{critt2})}}  \label{eq:entropy_functional} \\
&=&-\frac{\pi }{2}{\mathbf{v}}\mu _{\Lambda \Sigma }\left( \,\mathbf{v}%
\,\,e^{\Lambda }e^{\Sigma }+\frac{1}{\mathbf{v}}\,p^{\Lambda }p^{\Sigma
}\right) _{\text{(\ref{critt})-(\ref{critt2})}}.
\label{eq:entropy_functional_2}
\end{eqnarray}%
By recalling (\ref{Rel2}) and (\ref{eq:nearH2}), the result (\ref%
{eq:entropy_functional}) implies
\begin{equation}
\left. \mathcal{E}\right\vert _{\text{(\ref{critt})-(\ref{critt2})}}=S_{W}={%
\frac{\mathbf{v}}{4}}\left( e^{\Lambda }{\frac{\partial f}{\partial
e^{\Lambda }}}-f\right) _{\text{(\ref{critt})-(\ref{critt2})}}=\frac{S_{BH}}{%
4\pi }\left( e^{\Lambda }{\frac{\partial f}{\partial e^{\Lambda }}}-f\right)
_{\text{(\ref{critt})-(\ref{critt2})}},
\end{equation}%
which, by recalling that $S_{W}=S_{BH}$ (cf. (\ref{Rel1})) yields the
following result :%
\begin{equation}
\left( e^{\Lambda }{\frac{\partial f}{\partial e^{\Lambda }}}-f\right) _{%
\text{(\ref{critt})-(\ref{critt2})}}=4\pi ,  \label{tthis}
\end{equation}%
which, consistently, also follows from (\ref{This}) and (\ref{Rel2}).

\item In the formul\ae\ (\ref{This}), (\ref{eq:entropy_legendre}) and (\ref%
{Rel2}), the black hole entropy ($S_{W}$ or $S_{BH}$) of the dyonic extremal
black hole (\ref{eq:seed_solution}) depends on the magnetic ($p^{\Lambda }$)
and electric ($q_{\Lambda }$) charges of the black hole itself. Thus, in
such formul\ae\ we always understand the inversion of the relation %
\eqref{eq:qlambda_def} in terms of $e^{\Lambda }$ :
\begin{equation}
e^{\Lambda }=-{\frac{1}{\mathbf{v}}}\left[ \left( \mu ^{-1}\right) ^{\Lambda
\Sigma }\,\,q_{\Sigma }+(\mu ^{-1}\nu )_{\Sigma }^{\Lambda }\,\,p^{\Sigma }%
\right] .  \label{e-repl}
\end{equation}

\item It should be here remarked that, at the horizon, the equations of motion of scalar fields $z^a$ are given by the following criticality conditions (cf. \eqref{eq:sen_e_Vbh}) :
\begin{eqnarray}
{\frac{\delta \mathcal{E}}{\delta z^{a}}} =0= \frac{\delta V_{BH}}{\delta z^a}.
\label{critt2sc}
\end{eqnarray}
This solves the quite intriguing \emph{conundrum} of mismatching between different sets of scalar equations, discussed at the end of the last subsection. In fact, one can check that using the \emph{wrong potential} $\mathcal{V}_{BH}$ derived in \eqref{eq:wrong_pot}, one can write the criticality relations ${\delta \mathcal{V}_{BH}\over \delta z^a}={\delta f\over \delta z^a}=0$, which however does not yield the correct equations of motion for scalar fields near the horizon, given by \eqref{eq:eoms2}. In fact, as shown in \Cref{app:genRN}, \textit{it is not necessary to extremize the functional $f$ to compute the entropy}. The methodology underlying the construction of Sen's entropy functional may instead be viewed as a tailored construction of an effective action whose extremization reproduces the correct equations of motion.
\end{enumerate}

\noindent Summarizing, so far we have shown that%
\begin{equation}
\left. \mathcal{E}\right\vert _{\text{(\ref{critt}),(\ref{critt2}),(\ref%
{e-repl})}}=S_{W}=S_{BH}=\pi V_{BH,\text{hor}},  \label{eq:sen_e_Vbh}
\end{equation}%
where, from the discussion above, $V_{BH,\text{hor}}$ is the critical value
of $V_{BH}$ itself :
\begin{equation}
V_{BH,\text{hor}}=\left. V_{BH}\right\vert _{\text{(\ref{crit_cond})}}.
\end{equation}%
The formula (\ref{eq:sen_e_Vbh}) relates the critical value of Sen's entropy
functional $\mathcal{E}$ with the critical values of the effective black
hole potential $V_{BH}$ (yielding the Bekenstein-Hawking entropy $S_{BH}$),
and with the Wald black hole entropy $S_{W}$. of the black hole $S$.

\subsection{\label{R}The Routhian functional $\mathcal{R}$}

We will now proceed to show that the effective Routhian functional derived
from the full fledged Lagrangian density of the theory can also reproduce
the Wald entropy, as it is indeed easy to realize by observing that %
\eqref{eq:entropy_legendre} characterizes the Wald entropy $S_{W}$ as a
Legendre transform of $f$ (up to a total multiplicative factor). This also
relates the Routhian approach to the critical values of $\mathcal{E}$, see (%
\ref{eq:entropy_functional})-(\ref{eq:entropy_functional_2}).

We start from \eqref{ro} and express $\mathcal{R}$ in terms of $\partial
_{\tau }\psi ^{\Lambda }=e^{\Lambda }$ and $p^{\Lambda }$ only, rather than
writing it in the usual way in terms of $p^{\Lambda }$ and $q^{\Lambda }$ as
in \eqref{eq:finalrouth}. From \eqref{ro} and \eqref{eq:Routh_consts}, one
obtains that
\begin{equation}
-\frac{1}{2}e^{-2U}\left(
\begin{array}{cc}
\partial _{\tau }\psi ^{\Lambda }\ ~ & \partial _{\tau }\chi _{\Lambda }%
\end{array}%
\right)
\begin{pmatrix}
\left( \mu +\nu \mu ^{-1}\nu \right) _{\Lambda \Sigma } & -\left( \nu \mu
^{-1}\right) _{\Lambda }^{\Sigma } \\
&  \\
-\left( \mu ^{-1}\nu \right) _{\Sigma }^{\Lambda } & \left( \mu ^{-1}\right)
^{\Lambda \Sigma } \\
&
\end{pmatrix}%
\begin{pmatrix}
\partial _{\tau }\psi ^{\Sigma } \\
\\
\partial _{\tau }\chi _{\Sigma } \\
\end{pmatrix}%
=-{\frac{1}{2}}\left( e^{-2U}\mu _{\Lambda \Sigma }e^{\Lambda }e^{\Sigma
}+e^{2U}\mu _{\Lambda \Sigma }p^{\Lambda }p^{\Sigma }\right) ,
\end{equation}%
implying that the Routhian functional $\mathcal{R}$ \eqref{ro} can be
written as%
\begin{equation}
\mathcal{R}=\dot{U}^{2}+\frac{1}{2}G_{a\bar{a}}\dot{z}^{a}\dot{\bar{z}}^{%
\bar{a}}-{\frac{1}{2}}\mu _{\Lambda \Sigma }\left( e^{-2U}e^{\Lambda
}e^{\Sigma }+e^{2U}p^{\Lambda }p^{\Sigma }\right) .  \label{eq:routh_newform}
\end{equation}%
This expression is not manifestly invariant under (e.m.) duality, but rather
it is expressed only in terms of the symplectic-contravariant variables $%
e^{\Lambda }$ and $p^{\Lambda }$.

Next, in order to relate \eqref{eq:routh_newform} to the Wald entropy $S_{W}$
\eqref{eq:entropy_legendre}, we evaluate $\mathcal{R}$ at extremal black
hole horizon, and we also make a change of the radial coordinate from $\tau $
to $r:=1/\tau $, to match the current coordinate system. Therefore, the
effective action related to $\mathcal{R}$ (and denoted by $\mathfrak{S}$)
reads\footnote{%
While defining $\mathfrak{S}$, we already integrated out the $4\pi $ factor
that would come out of the angular integrals.}
\begin{equation}
\mathfrak{S}:=4\pi \int \mathcal{R}(\tau )\,\,d\tau =4\pi \int \mathcal{R}%
(r)\left( -{\frac{1}{r^{2}}}\right) \,dr=4\pi \int \mathcal{R}(r)\,dr.
\end{equation}%
The above transformation of the radial coordinate further implies that
\begin{equation}
e^{\Lambda }|_{\tau }=\partial _{\tau }\psi ^{\Lambda }=-\left( {\frac{1}{%
\tau ^{2}}}\right) \partial _{r}\psi ^{\Lambda }=-r^{2}e^{\Lambda }|_{r},
\end{equation}%
whereas the transformation of $p^{\Lambda }$ needs to be calculated more
carefully, because of the change in measure :
\begin{equation}
p^{\Lambda }|_{r}={\frac{\delta \mathcal{R}}{\delta (\partial _{r}\chi
_{\Lambda })}}={\frac{\delta \mathcal{R}(\tau )}{\delta (\partial _{\tau
}\chi _{\Lambda })}}=p^{\Lambda }|_{\tau }
\end{equation}%
All this jointly yields to the final result
\begin{equation}
\mathcal{R}_{\text{near-hor}}=-r^{2}{U^{\prime }}^{2}-{\frac{r^{2}}{2}}G_{a%
\bar{a}}\,z^{\prime a}\bar{z}^{\prime \bar{a}}+{\frac{1}{2}}\mu _{\Lambda
\Sigma }\left( \mathbf{v}\,e^{\Lambda }\,e^{\Sigma }+{\frac{1}{\mathbf{v}}}%
\,p^{\Lambda }\,p^{\Sigma }\right) ,  \label{eq:R_nearhorizon}
\end{equation}%
where "$\prime $" denotes the differentiation with respect to $r$. It is
here worth remarking that, \textit{exactly at the horizon}, both the first
and second term of (\ref{eq:R_nearhorizon}) vanish, because at the horizon $%
r=0$ and as $z^{\prime a}=0$ $\forall a$ (see the discussion above). Thus,
\begin{equation}
\mathcal{R}_{\text{hor}}={\frac{1}{2}}\left( \mathbf{v}\,\mu _{\Lambda
\Sigma }e^{\Lambda }\,e^{\Sigma }+{\frac{1}{\mathbf{v}}}\,\mu _{\Lambda
\Sigma }\,p^{\Lambda }\,p^{\Sigma }\right) _{\text{hor}}.
\label{eq:R_horizon}
\end{equation}
By multiplying \eqref{eq:R_horizon} $-\mathbf{v}\,\pi $, and recalling (\ref%
{eq:entropy_functional})-(\ref{eq:entropy_functional_2}) and (\ref{tthis}),
one finally achieves the following result :
\begin{equation}
-\mathbf{v}\,\pi \,\mathcal{R}_{\text{hor}}=\pi V_{BH,\text{hor}%
}=S_{BH}=\left. \mathcal{E}\right\vert _{\text{(\ref{critt})-(\ref{critt2})}%
}=S_{W}\ ,  \label{eq:final_circle}
\end{equation}%
which relates $S_{BH}=S_{W}$ to the Routhian functional $\mathcal{R}$
evaluated at the (unique) extremal black hole event horizon or,
equivalently, to Sen's entropy functional $\mathcal{E}$ evaluated at the
horizon or, again, to the black hole effective potential $V_{BH}$ at the
horizon. Another interesting way to show that $\mathcal{R}_{\text{hor}}$ would also
reproduce the Wald entropy $S_{W}$ through \eqref{eq:wald_Routhain} is the
following.

First, let us use the Hamiltonian constraint \eqref{eq:constraint} into %
\eqref{eq:static}, obtaining
\begin{equation}
\mathfrak{S}_{\text{on-shell}}=4\pi \int_{0}^{\infty }d\tau \ e^{2U(\tau
)}V_{BH}(\tau )\,.  \label{eq:R_density_def}
\end{equation}%
Then, upon changing $\tau =\frac{1}{r}$, we have
\begin{equation}
\mathfrak{S}_{\text{on-shell}}=4\pi \int \mathcal{R}\left( r\right) \,dr,~~%
\text{with~~}\mathcal{R}\left( r\right) =-\frac{1}{r^{2}}e^{2U(r)}V_{BH}(r)\
.  \label{eq:R_density_def2}
\end{equation}%
By using the relation given in \eqref{eq:at_horizon} (as $e^{2U}|_{\text{hor}%
}=\frac{r^{2}}{\mathbf{v}}$), we finally get
\begin{equation}
\mathcal{R}_{\text{hor}}=-\frac{1}{\mathbf{v}}V_{BH,\text{hor}},
\end{equation}%
thus proving \eqref{eq:final_circle}.

\section{Discussion and outlook}

\label{sec:Discussion}

We have provided a rigorous reformulation of the attractor dynamics at the
horizon of static, spherically symmetric and asymptotically flat extremal
black holes, solutions to the equations of motion of four-dimensional
Maxwell-Einstein-scalar theories : in this framework, we have shown that
the standard FGK effective action is naturally obtained as a \textit{%
Routhian reduction} of the underlying covariant theory, giving rise to the
relevant Routhian functional $\mathcal{R}$. This should be distinguished from the consistent one-dimensional reduction of \cite{deHaro:2003zd}, where the symmetry ansatz is imposed before varying the higher-dimensional action. In our case, since the charges are treated as the fixed physical quantities, the correct reduced action is rather obtained by a partial Legendre transform over the cyclic gauge variables, leading naturally to a Routhian reduction.

In this context, electric and
magnetic charges arise as conserved momenta, conjugate to \textit{cyclic}
gauge degrees of freedom, while the positivity of $V_{\mathrm{BH}}$ follows
from the suitable partial Legendre transform and, of course, from the
definiteness properties of the kinetic matrix $\mu _{\Lambda \Sigma }$. It
is also worth remarking that the Hamiltonian constraint, which is essential
to the entropy formula, appears as the residual manifestation of the
reparametrization invariance along the radial coordinate, on which the whole
effective dynamics is based.
\paragraph{The Routhian unification.}

A crucial conceptual payoff of the \textit{Routhian approach} is the natural
and elegant unification of the three major frameworks used in the literature
:

\begin{description}
\item[\textbf{i)}] the FGK formalism, based on the black hole (effective)
potential $V_{\mathrm{BH}}$ \cite{Ferrara:1997tw};

\item[\textbf{ii)}] Sen's entropy function method, based on the entropy
function(al) $\mathcal{E}$ \cite{Sen:2005wa,Sen:2006rmg};

\item[\textbf{iii)}] Wald's Noether-charge entropy prescription \cite%
{Wald:1993nt,Iyer:1994ys}.
\end{description}

We have shown that these frameworks are related by a web of partial Legendre
transforms implemented at different stages and with different choices of
variables; at the (unique) event horizon of the extremal black hole, the
criticality conditions of $V_{\mathrm{BH}}$, $\mathcal{E}$ and $\mathcal{R}$
all coincide, and they correctly reproduce the entropy. This provides a
structural explanation of why apparently distinct methods agree; at the same
time, this clarifies precisely where and why na\"{\i}ve reductions fail.

The result \eqref{eq:final_circle}, which also encompasses (\ref%
{eq:sen_e_Vbh}), closes a logical circle, like an \textit{ouroboros},
involving and relating all the relevant quantities discussed above :

\begin{itemize}
\item the on-shell effective $1$-dimensional action pertaining to the
Routhian functional $\mathcal{R}$ ($\mathfrak{S}_{\text{on-shell}}$);

\item the effective black hole potential at the horizon, i.e. at its
critical points ($V_{BH,\text{hor}}$);

\item Sen's entropy functional at its critical points ($\left. \mathcal{E}%
\right\vert _{\text{(\ref{critt})-(\ref{critt2})}}$);

\item the Bekenstein-Hawking entropy ($S_{BH}$);

\item the Wald entropy ($S_{W}$).
\end{itemize}

Our results also fit naturally into a broader modern landscape of attractor
physics. Non-supersymmetric attractors \cite{Tripathy:2005qp} and
first-order flow descriptions based on fake superpotentials \cite%
{Ceresole:2007wx,Bossard:2009we} can be conceived as different
parameterizations of the same reduced phase-space dynamics, suggesting that
Routhian methods may offer a clean way to organize (and perhaps classify)
admissible first-order formulations. Furthermore, the symplectic and
manifestly duality covariant structure of $\mathcal{N}=2,D=4$ supergravity
\cite{Ceresole:1995ca,Andrianopoli:1996cm} is naturally consistent with the
Routhian formalism, and it hints at a systematic incorporation of
duality-invariant operations on charges, including Freudenthal duality (and
generalizations thereof) \cite{Borsten:2009zy,Borsten:2012pd,Ferrara:2011gv}.\bigskip

For what concerns an outlook to the future, several future directions are
particularly promising.

\paragraph{Higher-derivative interactions and generalized attractors.}

Since Sen's entropy function and Wald entropy remain applicable in the
presence of higher-derivative terms (as it is the case of effective
Lagrangians from string theory), a Routhian-based derivation of the
effective dynamics may provide a systematic extension of the attractor flows
beyond two derivatives. This could help clarify the relation between
higher-derivative corrected potentials, stability conditions, and entropy
extremization, relating to early studies of attractors in higher-curvature
theories \cite{Chandrasekhar:2006kx}. Notwithstanding the fact that in \cite%
{Bellucci:2009nn} it has been argued that the entropy function formalism is
more suitable than the FGK formalism standard attractors in the context of
higher-derivative corrections, we here take the chance to anticipate that
the inter-connection between these two approaches is likely to be way more
intricate than what \cite{Bellucci:2009nn} stated.

\paragraph{Duality-covariant reduction and geometric mechanics.}

In geometric mechanics, the Routhian formalism is naturally interpreted as
symmetry reduction. For this reason, the development and investigation of a
fully geometric version of the present construction - in which the
symplectic bundle of special K\"{a}hler geometry and the electric--magnetic
duality group act as reduction data - can be regarded as particularly
desirable. Such a geometric formulation may sharpen the understanding of how
duality covariance is maintained off-shell, as well as how reduced
\textquotedblleft magnetic\textquotedblright\ terms may arise in more
general settings.

\paragraph{Beyond static spherical symmetry.}

Extensions to stationary (rotating), asymptotically non-flat, multi-center,
topological solutions and less symmetric horizons remain an important and
challenging frontier. Given the close relation between first-order flows,
stability and multi-center physics \cite{Denef:2000nb}, it is natural to ask
whether Routhian techniques can provide a clean variational basis for these
generalizations, and how they interface with near-horizon $\mathrm{AdS}_{2}$
symmetry and with $\mathrm{AdS}_{2}/\mathrm{CFT}_{1}$ ideas.

\paragraph{Relation to alternative variables and generalized flow formalisms.%
}

The H-FGK formalism, introduced and developed in \cite%
{Meessen:2011aa,Meessen:2012vh}, suggests that suitably chosen variables can
dramatically simplify the construction of both extremal and non-extremal
solutions. It would be worthwhile to explore how such variable choices
appear withon the Routhian formalism, and whether this latter provides a
systematic route to identifying \textquotedblleft good\textquotedblright\
variables adapted to duality and integrability.

\paragraph{(Generalized) Freudenthal duality in the Routhian formalism?}

We argue that (\ref{eq:R_nearhorizon}) is the correct off-the-horizon
generalisation of the Wald entropy. This fact hints at the tantalizing
possibility to generalize Freudenthal duality (or a scalar dependent version
thereof \cite{Borsten:2012pd,Ferrara:2011gv}) within the Routhian
formalism.\bigskip

We expect that the Routhian unification developed in this paper will provide
a sound, conceptual foundation for the above developments, while providing a
clean and elegant bridge between effective-dynamics, near-horizon, and
duality-covariant approaches to extremal black hole physics.

\section*{Acknowledgments}

We would like to thank Kostas Skenderis (University of Southampton), P. N. Bala Subramanian (NIT Calicut) and Taniya Mandal (NISER) for illuminating discussions.

SRC would also like to thank the organizers and participants of {\it{Shanghai-APCTP-Sogang-GIST workshop on Gravity, Astroparticle, and Cosmology (SASGAC 2025)}} for an opportunity to present a part of this work and for their important feedback. 
The work of SRC is supported in part by the \textit{National Research
Foundation of Korea} (NRF) grant funded by the Korea government (MSIT) with
grant number RS-2024-00449284, the Sogang University Research Grant of
202410008.01, the Basic Science Research Program of the \textit{National
Research Foundation of Korea} (NRF) funded by the Ministry of Education
through the \textit{Center for Quantum Spacetime} (CQUeST) with grant number
RS-2020-NR049598. 
\appendix

\section{All other, alternative, possible Routhians yield to incorrect
potentials !}

\label{app:others}
In our prescription starting from \cref{subsec:pot_routh}, we first introduce $\psi^\Lambda$. without introducing the dual field strength $\mathcal{G}_{\Lambda }$. In this appendix we explore that other case, where $\psi^\Lambda$ and $\chi_\Lambda$ are both introduced and we have \eqref{eq:eff_lag_4d} as 
\begin{eqnarray}
L &=&-\sin \theta \left( \dot{U}^{2}+\frac{1}{2}G_{a\bar{a}}\dot{z}^{a}\dot{%
\bar{z}}^{\bar{a}}\right) +\frac{1}{2}e^{-2U}\sin \theta \ \left(
\begin{array}{cc}
\partial _{\tau }\psi ^{\Lambda }\ ~ & \partial _{\tau }\chi _{\Lambda }%
\end{array}%
\right)
\begin{pmatrix}
-\left( \mu +\nu \mu ^{-1}\nu \right) _{\Lambda \Sigma } & \left( \nu \mu
^{-1}\right) _{\Lambda }^{\Sigma } \\
&  \\
-\left( \mu ^{-1}\nu \right) _{\Sigma }^{\Lambda } & \left( \mu ^{-1}\right)
^{\Lambda \Sigma } \\
&
\end{pmatrix}%
\begin{pmatrix}
\partial _{\tau }\psi ^{\Sigma } \\
\\
\partial _{\tau }\chi _{\Sigma } \\
\end{pmatrix}%
\ ,  \notag  \label{for vbhcon42 totop} \\
&&
\end{eqnarray}
In this case, we have two cyclic coordinates, and therefore have three choices for the Routhian, by taking partial Legendre transformation with respect to: \textbf{i}) $\psi ^{\Lambda }$; \textbf{ii}) $\chi _{\Lambda }$; or \textbf{%
iii}) both of them denoted respectively as $\mathcal{R}_{\psi}$, $\mathcal{R}_{\chi}$ and $\mathcal{R}_{\psi,\chi}$.

\begin{description}
\item[i) $\mathcal{R}_{\psi}$:] As evident from before, for this case we can define
\begin{align}
\mathcal{R}_{ \psi }:=\partial _{\tau
}\psi^{\Lambda }{\frac{\delta \mathcal{L}}{\delta (\partial _{\tau }\psi
^{\Lambda })}}-\mathcal{L}\ ,
\end{align}
which leads to (upto an overall factor of $4\pi$) as
\begin{align}
\mathcal{R}_{\psi } &=&\dot{U}^{2}+\frac{1}{2}G_{a\bar{a}}\dot{z}^{a}\dot{%
\bar{z}}^{\bar{a}}-\frac{1}{2}e^{-2U}\left(
\begin{array}{cc}
\partial _{\tau }\psi ^{\Lambda }\ ~ & \partial _{\tau }\chi _{\Lambda }%
\end{array}%
\right)
\begin{pmatrix}
\left( \mu +\nu \mu ^{-1}\nu \right) _{\Lambda \Sigma } & 0 \\
&  \\
0 & \left( \mu ^{-1}\right) ^{\Lambda \Sigma } \\
&
\end{pmatrix}%
\begin{pmatrix}
\partial _{\tau }\psi ^{\Sigma } \\
\\
\partial _{\tau }\chi _{\Sigma } \\
\end{pmatrix}%
\ ,  \notag\\
&&
\end{align}%
\item[ii) $\mathcal{R}_{\chi}$:] With the same spirit as above, here we have (upto an overall factor of $4\pi$)
\begin{eqnarray}
\mathcal{R}_{\chi } &=&\dot{U}^{2}+\frac{1}{2}G_{a\bar{a}}\dot{z}^{a}\dot{%
\bar{z}}^{\bar{a}}-\frac{1}{2}e^{-2U}\left(
\begin{array}{cc}
\partial _{\tau }\psi ^{\Lambda }\ ~ & \partial _{\tau }\chi _{\Lambda }%
\end{array}%
\right)
\begin{pmatrix}
-\left( \mu +\nu \mu ^{-1}\nu \right) _{\Lambda \Sigma } & 0 \\
&  \\
0 & -\left( \mu ^{-1}\right) ^{\Lambda \Sigma } \\
&
\end{pmatrix}%
\begin{pmatrix}
\partial _{\tau }\psi ^{\Sigma } \\
\\
\partial _{\tau }\chi _{\Sigma } \\
\end{pmatrix}%
\ ,  \notag \\
&&
\end{eqnarray}
\item[iii) $\mathcal{R}_{\psi,\chi}$:] For this case we can define,
\begin{align}
\mathcal{R}_{ \psi,\chi }:=\partial _{\tau
}\psi^{\Lambda }{\frac{\delta \mathcal{L}}{\delta (\partial _{\tau }\psi
^{\Lambda })}}+\partial _{\tau
}\chi _{\Lambda }{\frac{\delta \mathcal{L}}{\delta (\partial _{\tau }\chi
_{\Lambda })}}-\mathcal{L}\ ,
\end{align}
which results into (upto an overall factor of $4\pi$)
\begin{eqnarray}
\mathcal{R}_{\psi,\chi } &=&\dot{U}^{2}+\frac{1}{2}G_{a\bar{a}}\dot{z}^{a}\dot{%
\bar{z}}^{\bar{a}}-\frac{1}{2}e^{-2U}\left(
\begin{array}{cc}
\partial _{\tau }\psi ^{\Lambda }\ ~ & \partial _{\tau }\chi _{\Lambda }%
\end{array}%
\right)
\begin{pmatrix}
-\left( \mu +\nu \mu ^{-1}\nu \right) _{\Lambda \Sigma } & \left( \nu \mu
^{-1}\right) _{\Lambda }^{\Sigma } \\
&  \\
-\left( \mu ^{-1}\nu \right) _{\Sigma }^{\Lambda } & \left( \mu ^{-1}\right)
^{\Lambda \Sigma } \\
&
\end{pmatrix}
\begin{pmatrix}
\partial _{\tau }\psi ^{\Sigma } \\
\\
\partial _{\tau }\chi _{\Sigma } \\
\end{pmatrix}%
\ ,  \notag \\
&&
\end{eqnarray}
\end{description}
None of the three cases reproduces the correct form of $V_{BH}$ as in \eqref{eq:normal_effective}.

\section{Alternative derivation of $V_{BH}~$: the road not (\textit{to be})
taken}

\label{app:noroad}

This appendix focusses on the procedure carried out by Kallosh, Sivanandam
and Soroush (KSS) in \cite{Kallosh:2006bt} in order to derive the correct
effective action starting from \eqref{eq:lag_eff}, instead of constructing
the Routhian functional defined in \eqref{ro}.

KSS\ start and introduce the complex dual field strengths
\begin{equation}
\mathcal{G}_{\Lambda |\mu \nu }:=-i\mu _{\Lambda \Sigma }({\star }F)_{\mu
\nu }^{\Sigma }+\nu _{\Lambda \Sigma }F_{\mu \nu }^{\Sigma }\ ,
\label{im du}
\end{equation}%
along with defining magnetic potential $\chi _{\Lambda }$ as%
\begin{equation}
\mathcal{G}_{\Lambda \tau t}=:\partial _{\tau }\chi _{\Lambda },
\end{equation}%
thus obtaining the purely imaginary field strength
\begin{equation}
\sqrt{-g}F^{\Lambda \theta \phi }=i(\mu ^{-1})^{\Lambda \Sigma }\partial
_{\tau }\chi _{\Sigma }-i(\mu ^{-1}\nu )_{\Sigma }^{\Lambda }\partial _{\tau
}\psi ^{\Sigma },  \label{im du2}
\end{equation}%
where%
\begin{equation}
F_{\tau t}^{\Lambda }=:\partial _{\tau }\psi ^{\Lambda }.
\end{equation}

By plugging \eqref{im du2} into \eqref{eq:lag_eff} and considering only the
real part, one obtains the expression of the same effective Lagrangian as
the one given by the r.h.s. of \eqref{ro}. On the other hand, it should be
remarked that, upon plugging \eqref{im du2} into \eqref{eq:lag_eff} and
setting the imaginary part to zero, the following constraint is obtained :
\begin{equation}
\nu _{\Lambda \Sigma }F_{\tau t}^{\Lambda }F^{\Sigma \theta \phi
}=0\Longrightarrow \left( \nu \mu ^{-1}\right) _{\Lambda }^{\Gamma }\partial
_{\tau }\psi ^{\Lambda }\partial _{\tau }\chi _{\Gamma }-\left( \nu \mu
^{-1}\nu \right) _{\Lambda \Gamma }\partial _{\tau }\psi ^{\Lambda }\partial
_{\tau }\psi ^{\Gamma }=0\ .  \label{im 0}
\end{equation}

The whole KSS procedure \cite{Kallosh:2006bt} relied on defining the \textit{%
complex} dual field strengths exactly as in \eqref{im du}, without
justifying it in any way. Also, KSS neglected the constraint (\ref{im 0}),
which is indeed unnecessary, and thus looks unphysical. In order to avoid
the unjustified definition (\ref{im du}) of complex dual field stregthts as
well as the resulting unphysical constraint (\ref{im 0}), we put forward our
Routhian prescription, based on the definition of an effective Routhian
functional through a partial Legendre transform : indeed, Routhian formalism
provides a rigorous framework which produces and explains the correct
effective 1-dimensional action \eqref{eq:normal_effective}, originally
guessed by Ferrara, Gibbons and Kallosh in their seminal work \cite%
{Ferrara:1997tw} by inspecting the resulting equations of motion.

\section{Entropy functional with $SO(2,1)\times SO(3)$ isometry : Sen's
derivation}

\label{app:genRN}

In this appendix, we will derive Sen's entropy functional $\mathcal{\tilde{E}%
}$ for the most general static, spherically symmetric and asymptotically
flat extremal black hole in four space-time dimensions, by focusing on its
near-horizon region with $SO(2,1)\times SO(3)$ isometry, whose most general
metric is given by\footnote{%
Note that we are not imposing $\mathbf{v}_{1}=\mathbf{v}_{2}$ \textit{a
priori}, and thus (\ref{a1}) is not the Bertotti-Robinson metric (but it
will become so, after the criticality conditions of $\tilde{f}$ with respect
to $\mathbf{v}_{1}$ and $\mathbf{v}_{2}$ (\ref{eq:vrel}) will be taken into
account). }\footnote{Note a sign change in metric from mostly-minus to mostly-plus, to make the results consistent with the literature} \cite{Sen:2007qy,Astefanesei:2006dd}
\begin{equation}
ds^{2}=\mathbf{v}_{1}\left( -r^{2}dt^{2}+\frac{dr^{2}}{r^{2}}\right) +%
\mathbf{v}_{2}\left( d\theta ^{2}+\sin ^{2}\theta d\phi ^{2}\right) ,
\label{a1}
\end{equation}%
along with the background fluxes $F_{rt}^{\Lambda }=e^{\Lambda }$ and $%
F_{\theta \phi }^{\Lambda }=-p^{\Lambda }\sin \theta $. The radii $\mathbf{v}%
_{1}$ and $\mathbf{v}_{2}$, as well as $\{e^{\Lambda }\}$, $\{p^{\Lambda }\}$
and the scalar fields $z^{a}$, take constant values when approaching the
(unique) event horizon of the extremal black hole, since the attractor
mechanism is at work
. We should stress that, with respect
to the treatment of Sec. \eqref{sec:efd}, in this appendix we will define tilded
functionals (such as $\tilde{f}$ and $\mathcal{\tilde{E}}$) which are
slightly more general than the untilded ones ($f$ and $\mathcal{E}$,
respectively treated in Secs. \ref{f} and \ref{E}), but which, as we will
see further below, match the latter ones after the criticality conditions (%
\ref{eq:vrel}) of $\tilde{f}$ with respect to $\mathbf{v}_{1}$ and $\mathbf{v%
}_{2}$ have been taken into account.

By following the same \textit{rationale} underlying the treatment given in
section \eqref{sec:efd}, we start and define the functional $\tilde{f}(z^{a},%
\bar{z}^{\bar{a}},\mathbf{v}_{i},e^{\Lambda },p^{\Lambda })$ ($i=1,2$) as
the Lagrangian density $L$ integrated over the angular coordinates in the
near-horizon limit (\ref{a1}) 
\begin{eqnarray}
\tilde{f}(z^{a},\bar{z}^{\bar{a}},\mathbf{v}_{i},e^{\Lambda },p^{\Lambda })
&:&=\int d\theta d\phi \ \sqrt{-g}\left. L\right\vert _{\text{(\ref{a1})}%
}=4\pi (\mathbf{v}_{2}-\mathbf{v}_{1})-2\pi \,\frac{\mathbf{v}_{2}}{%
\mathbf{v}_{1}}\,\mu _{\Lambda \Sigma }\,\left( z^{a},\bar{z}^{\bar{a}%
}\right) \,e^{\Lambda }e^{\Sigma }  \notag \\
&&+2\pi \frac{\mathbf{v}_{1}}{\mathbf{v}_{2}}\mu _{\Lambda \Sigma
}\,\left( z^{a},\bar{z}^{\bar{a}}\right) \,p^{\Lambda }p^{\Sigma }-4\pi
\,\nu _{\Lambda \Sigma }\,\left( z^{a},\bar{z}^{\bar{a}}\right) \,e^{\Lambda
}p^{\Sigma }\ .  \label{a2}
\end{eqnarray}%
The differentiating with respect to $e^{\Lambda }$ entails the relations
\begin{equation}
4\pi q_{\Lambda }:=\frac{\partial \tilde{f}}{\partial e^{\Lambda }}=-4\pi
\left( \frac{\mathbf{v}_{2}}{\mathbf{v}_{1}}\ \mu _{\Lambda \Sigma
}e^{\Sigma }+\nu _{\Lambda \Sigma }p^{\Sigma }\right) \ ,
\label{eq:e_qrelation}
\end{equation}%
in which the black hole electric charges $q_{\Lambda }$ are defined as the
\textit{conjugates} of $e^{\Lambda }$.

The definition of $q_{\Lambda }$ is instrumental to the introduction of a
new functional $\mathcal{\tilde{E}}(z^{a},\bar{z}^{\bar{a}},\mathbf{v}%
_{i},e^{\Lambda },p^{\Lambda },q_{\Lambda })$ as
\begin{equation}
\mathcal{\tilde{E}}(z^{a},\bar{z}^{\bar{a}},\mathbf{v}_{i},e^{\Lambda
},p^{\Lambda },q_{\Lambda }):=\widetilde{\mathfrak{n}}\left( 4\pi e^{\Lambda
}q_{\Lambda }-\tilde{f}(z^{a},\bar{z}^{\bar{a}},\mathbf{v}_{i},e^{\Lambda
},p^{\Lambda })\right) ,\label{E-tilde}
\end{equation}%
namely as the Legendre transform of the functional $\tilde{f}$ with respect
to the electric charge $q_{\Lambda }$. Note that $\widetilde{\mathfrak{n}}$
is a constant factor which plays a key role, as we will see below.
Remarkably, the extremization of $\mathcal{\tilde{E}}(z^{a},\bar{z}^{\bar{a}%
},\mathbf{v}_{i},e^{\Lambda },p^{\Lambda },q_{\Lambda })$ allows one to
recover \textit{all} the near-horizon equations of motion
; furthermore, the
overall factor $\widetilde{\mathfrak{n}}$ can be chosen such that the
critical values of $\mathcal{\tilde{E}}$ compute the Wald entropy $S_{W}$.
Following \eqref{sec:efd}, one can check that, for the metric \eqref{a1},
\begin{equation}
S_{W}=-\frac{1}{4}\left. \frac{\partial
\tilde{f}_{\lambda }}{\partial \lambda }\right\vert _{\lambda =1},
\end{equation}%
implying%
\begin{equation}
\widetilde{\mathfrak{n}}=\frac{1}{4},
\end{equation}%
which allows one to recast the formula (\ref{E-tilde}) of the entropy
function as follows\footnote{%
Notice the difference in constant factors between the definition due to
Ashoke Sen in \cite{Sen:2007qy}, following which $\mathcal{E}=2\pi
(e_{i}q_{i}-f)$. This is because of the difference in the overall factor of %
\eqref{eq:theaction} and the use of rationalised Heaviside-Lorentz unit in
our case where $\epsilon _{0}=1$, bringing in the extra factor of $4\pi $ in
the definition of $\mathcal{E}$.} :
\begin{equation}
\mathcal{\tilde{E}}(z^{a},\bar{z}^{\bar{a}},\mathbf{v}_{i},e^{\Lambda
},p^{\Lambda },q_{\Lambda })=\frac{1}{4}%
\left( 4\pi e^{\Lambda }q_{\Lambda }-\hat{f}(z^{a},\bar{z}^{\bar{a}},\mathbf{%
v}_{i},e^{\Lambda },p^{\Lambda })\right) .  \label{eq:Edef}
\end{equation}

The remarkable feature of Sen's entropy function(al) $\mathcal{\tilde{E}}%
(z^{a},\bar{z}^{\bar{a}},\mathbf{v}_{i},e^{\Lambda },p^{\Lambda },q_{\Lambda
})$ is that its criticality conditions provide algebraic equations whose
solution directly yields the entropy of the black hole under consideration.
In this case, we can start by
\begin{equation}
{\frac{\partial {\mathcal{\tilde{E}}}}{\partial e^{\Lambda }}}=0\implies
q_{\Lambda }=-\left( \frac{\mathbf{v}_{2}}{\mathbf{v}_{1}}\ \mu
_{\Lambda \Sigma }e^{\Sigma }+\nu _{\Lambda \Sigma }p^{\Sigma }\right)
\label{eq:eqagain}
\end{equation}%
where we have used \eqref{eq:e_qrelation}. Since we aim at writing the black
hole entropy in terms of the electric charges $q_{\Lambda }$ and of the
magnetic charges $p^{\Lambda }$, the relations (\ref{eq:eqagain}) can be
solved in terms of $e^{\Lambda }$ (by exploiting the invertibility of the
kinetic vector matrix $\mu _{\Lambda \Sigma }$), obtaining
\begin{equation}
e^{\Lambda }=-\frac{\mathbf{v}_{1}}{\mathbf{v}_{2}}\left[ \left( \mu
^{-1}\right) ^{\Lambda \Sigma }q_{\Sigma }+(\mu ^{-1}\nu )_{\Sigma
}^{\Lambda }p^{\Sigma }\right] \ .  \label{a3}
\end{equation}

By plugging (\ref{a3}) and the expression of $\tilde{f}(z^{a},\bar{z}^{\bar{a%
}},\mathbf{v}_{i},e^{\Lambda },p^{\Lambda })$ in \eqref{a2} into the functional\\
$\mathcal{\tilde{E}}(z^{a},\bar{z}^{\bar{a}},\mathbf{v}%
_{i},e^{\Lambda },p^{\Lambda },q_{\Lambda })$ in \eqref{eq:Edef}, one
obtains
\begin{equation}
\left. \mathcal{\tilde{E}}(z^{a},\bar{z}^{\bar{a}},\mathbf{v}_{i},e^{\Lambda
},p^{\Lambda },q_{\Lambda })\right\vert _{\text{(\ref{a3})}}=\pi \left[ (%
\mathbf{v}_{1}-\mathbf{v}_{2})+\frac{\mathbf{v}_{1}}{\mathbf{v}_{2}}%
V_{BH}\left( z^{a},\bar{z}^{\bar{a}},p^{\Lambda },q_{\Lambda }\right) \right]
\ .  \label{eq:Edef2}
\end{equation}%
This expression provides a relation between Sen's entropy functional $%
\mathcal{\tilde{E}}$, introduced by Sen in \cite{Sen:2005wa}, and the $1$%
-dimensional effective black hole potential $V_{BH}$, introduced by FGK in
\cite{Ferrara:1997tw}. To the best of our knowledge, this relation was
firstly dicussed by Trivedi \cite{Trivedi2007} (by exploiting the criticality
condition \eqref{eq:eqagain}).

On the other hand, from (\ref%
{eq:Edef2}) the criticality conditions of $\mathcal{\tilde{E}}$ with
respect to $z^{a}$,  are equivalent to the criticality conditions of $V_{BH}$ with
respect to $z^{a}$, imply the equations of motion of the $z^{a}$'s
themselves, and thus, in the near-horizon region, their stabilization (into
purely black hole charge dependent values) determined by the attractor
mechanism :%
\begin{equation}
\frac{\partial \mathcal{\tilde{E}}}{\partial z^{a}}=0%
\overset{\text{(\ref{eq:Edef2})}}{\Leftrightarrow }\frac{\partial V_{BH}}{%
\partial z^{a}}=0.
\end{equation}%
Moreover, the criticality conditions of $\mathcal{\tilde{E}}$ with respect to $\mathbf{%
v}_{1}$ and $\mathbf{v}_{2}$ imply such two radii to be equal,
\begin{equation}
{\frac{\partial \mathcal{\tilde{E}}}{\partial \mathbf{v}_{1}}}=0={\frac{\partial \mathcal{\tilde{E}}%
}{\partial \mathbf{v}_{2}}}\implies \mathbf{v}_{1}=\mathbf{v}_{2}=:\mathbf{v}%
,  \label{eq:vrel}
\end{equation}%
thus transforming the $SO(2,1)\times SO(3)$ covariant near-horizon metric (%
\ref{a1}) into the Bertotti-Robinson one, given by (\ref{eq:nearH}). Also,
by implementing the criticality conditions (\ref{eq:vrel}) and recalling the
treatment of Sec. \ref{sec:efd}, one obtains that%
\begin{eqnarray}
\left. \tilde{f}(z^{a},\bar{z}^{\bar{a}},\mathbf{v}_{1},\mathbf{v}%
_{2},e^{\Lambda },p^{\Lambda })\right\vert _{\text{(\ref{eq:vrel})}}
&=&f\left( z^{a},\bar{z}^{\bar{a}},e^{\Lambda },p^{\Lambda }\right) ; \\
\left. \mathcal{\tilde{E}}(z^{a},\bar{z}^{\bar{a}},\mathbf{v}_{1},\mathbf{v}%
_{2},e^{\Lambda },p^{\Lambda })\right\vert _{\text{(\ref{eq:vrel})}} &=&%
\mathcal{E}\left( z^{a},\bar{z}^{\bar{a}},e^{\Lambda },p^{\Lambda }\right) ;
\\
\left. \mathcal{\tilde{E}}(z^{a},\bar{z}^{\bar{a}},\mathbf{v}_{i},e^{\Lambda
},p^{\Lambda },q_{\Lambda })\right\vert _{\text{(\ref{a3}),(\ref{eq:vrel})}}
&=&\left. \mathcal{E}\left( z^{a},\bar{z}^{\bar{a}},e^{\Lambda },p^{\Lambda
}\right) \right\vert _{\text{(\ref{a3})}}=\pi V_{BH}\left( z^{a},\bar{z}^{%
\bar{a}},p^{\Lambda },q_{\Lambda }\right) ; \\
\left. \widetilde{\mathfrak{n}}\right\vert _{\text{(\ref{eq:vrel})}} &=&%
\frac{1}{4}\overset{\text{(\ref{Rel2})}}{=}\mathfrak{n}.
\end{eqnarray}%
Finally, by taking into account all results from \eqref{eq:Edef}, %
\eqref{eq:eqagain}, \eqref{a3}, \eqref{eq:Edef2} and \eqref{eq:vrel}, one
can then evaluate the extremum (i.e., critical) value of Sen's entropy
functional $\mathcal{\tilde{E}}$ as
\begin{eqnarray}
\left. \mathcal{\tilde{E}}(z^{a},\bar{z}^{\bar{a}},\mathbf{v}_{1},\mathbf{v}%
_{2},e^{\Lambda },p^{\Lambda })\right\vert _{\text{extremized}} &=&\left.
\mathcal{E}(z^{a},\bar{z}^{\bar{a}},e^{\Lambda },p^{\Lambda })\right\vert _{%
\text{extremized}} \\
&=&\pi \left. V_{BH}\right\vert _{\frac{\partial V_{BH}}{\partial z^{a}}%
=0}=\pi V_{BH,\text{hor}}=S_{BH} \\
&=&\frac{\mathbf{1}}{4}\left( e^{\Lambda }{\frac{\partial f}{\partial
e^{\Lambda }}}-f\right) _{\text{(\ref{critt})-(\ref{critt2})}}=S_{W}.
\end{eqnarray}

\noindent{It should be stressed that \textit{in none of the above derivations one needs to extremize the functional $f$}.}

\section{\label{App-Example-Routh}An application of the Routhian formalism :
central-force reduction and effective potential}

A canonical textbook example (going back to the classical treatises, such as
\cite{Whittaker_1988}) is the so-called `planar central-force problem' (see
also e.g. \cite{Goldstein2001}).

In polar coordinates $(r,\theta )$ on $\mathbb{R}^{2}$ with potential $V(r)$%
,
\begin{equation}
\mathbf{L}(r,\theta ,\dot{r},\dot{\theta})=\frac{m}{2}(\dot{r}^{2}+r^{2}\dot{%
\theta}^{2})-V(r).  \label{eq:Lcentral}
\end{equation}%
The coordinate $\theta $ is \textit{cyclic}, so
\begin{equation}
p_{\theta }:=\frac{\delta L}{\delta \dot{\theta}}=mr^{2}\dot{\theta}=\ell
\label{eq:angmom}
\end{equation}%
(angular momentum) is conserved in time :%
\begin{equation}
\dot{p}_{\theta }=0.  \label{cons}
\end{equation}%
Solving (\ref{eq:angmom}) for $\dot{\theta}$ yields $\dot{\theta}=\ell
/(mr^{2})$, and therefore the Routhian functional $\mathcal{R}$ %
\eqref{eq:Rdef} in this case reads
\begin{align}
\mathcal{R}(r,\dot{r},\ell )& =\left[ \frac{m}{2}\dot{r}^{2}+\frac{m}{2}r^{2}%
\dot{\theta}^{2}-V(r)\right] -\ell \dot{\theta}  \notag \\
& =\frac{m}{2}\dot{r}^{2}-V(r)-\frac{\ell ^{2}}{2mr^{2}}.
\label{eq:Rcentral}
\end{align}

Thus, the \textit{reduced} variational principle is governed by the purely
radial Lagrangian functional
\begin{equation}
\mathbf{L}_{\ell }(r,\dot{r})=\frac{m}{2}\dot{r}^{2}-\left( V(r)+\frac{\ell
^{2}}{2mr^{2}}\right) ,  \label{eq:Lell}
\end{equation}%
whose Euler--Lagrange equation gives the radial dynamics in the effective
potential
\begin{equation}
V_{\text{eff}}(r):=V(r)+\frac{\ell ^{2}}{2mr^{2}},
\end{equation}%
the so-called \textquotedblleft centrifugal barrier\textquotedblright\ \cite%
{Goldstein2001}.

Finally, the reconstruction follows from \eqref{eq:ydot_from_R}:
\begin{equation}
\dot{\theta}=-\frac{\partial \mathcal{R}}{\partial \ell }=\frac{\ell }{mr^{2}%
},
\end{equation}%
which allows to recover the conserved angular momentum law.

This example highlights a general mechanism : \textit{eliminating a cyclic
variable trades a second-order equation for a first-order constraint, and
correspondingly modifies the reduced Lagrangian functional by a
\textquotedblleft centrifugal\textquotedblright , momentum-dependent term}.

\section{\label{App-Magn}Geometric Routh reduction and `magnetic'\ terms}

The coordinate construction above can be regarded as the Abelian shadow of a
more general and deep reduction procedure, which we briefly report here; for
further details, we address the interested reader e.g. to \cite%
{MarsdenMontgomeryRatiu1990,MarsdenRatiu1999,HolmSchmahStoica2009}.

Let a Lie group\footnote{%
For general, non-Abelian $G$, cyclic coordinates rarely exist globally. In
the framework pertaining to the present investigation, $G=U(1)^{n+1}$ is the
group of the gauge symmetry, where $n+1$ is the number of Abelian 1-form
potentials.} $G$ act freely and properly on $Q$, and let us assume $\mathbf{L%
}$ is $G$--invariant\footnote{%
The Lie algebras of $G$ and its dual are denoted by $\mathfrak{g}$ and $%
\mathfrak{g}^{\ast }$, respectively.}. Noether's theorem produces a
conserved \textit{momentum map}
\begin{equation}
J\colon TQ\rightarrow \mathfrak{g}^{\ast },
\end{equation}%
and for fixed $\mu \in \mathfrak{g}^{\ast }$ one considers the \textit{%
momentum level} $J^{-1}(\mu )$. Then, the \textit{Routh reduction} can
briefly be described as a 3-steps procedure :

\begin{enumerate}
\item restrict the dynamics to $J^{-1}(\mu )$;

\item quotient by the isotropy subgroup $G_{\mu }$ (or by $G$ when Abelian);

\item obtain reduced equations on a space fibering over the `\textit{shape
space}' $Q/G$ \cite{MarsdenRatiu1999,HolmSchmahStoica2009}. In ther words,
one obtains a \textit{reduced} variational principle on (a bundle over) $Q/G$%
.
\end{enumerate}

Therefore, the \textit{Routhian reduction} exploits a connection to separate
vertical/horizontal velocities, hence yielding \textit{reduced} equations of
motion which are the Euler--Lagrange equations on $Q/G$, coupled to \textit{%
reconstruction} equations on the group $G$ itself.

A technical subtlety, which is invisible in the global cyclic-coordinate
case, is that the separation of \textquotedblleft group\textquotedblright\
and \textquotedblleft shape\textquotedblright\ velocities requires a
principal connection on $Q\rightarrow Q/G$. In local trivializations, the%
\textit{\ reduced} Lagrangian resembles \eqref{eq:Lmu}, but it may acquire
additional \textit{gyroscopic} (or \textit{`magnetic'}) terms, containing a
momentum-dependent one-form on $T(Q/G)$ whose exterior derivative acts as an
effective `\textit{curvature force}' \cite%
{MarsdenMontgomeryRatiu1990,HolmSchmahStoica2009} (this remark clarifies why
reduced equations may contain Lorentz-type forces even when the original
Lagrangian is of simple kinetic-minus-potential type). However, it is here
wprth remarking that in the Abelian/cyclic-coordinate(s') case, these terms
vanish, and one recovers the coordinate Routhian functional \eqref{eq:Rdef}.

Remarkably, Routh reduction is relevant \textit{per se} within the broader
framework in which \textquotedblleft reduction by stages\textquotedblright\
is carried out (consisting of stepwise symmetry reductions, possible when
the symmetry group $G$ has normal subgroups), which systematically organizes
partial reductions closely related to Routhians \cite{MarsdenEtAl2007}.


\bibliographystyle{jhep}
\bibliography{T_bib}
{}

\end{document}